\documentstyle[12pt]{article}
\begin{document}
\author{\normalsize\bf Yu.A. Markov \thanks{e-mail:markov@icc.ru}
and M.A. Markova$^*$}
\title{The Boltzmann equation for colorless plasmons\\
in hot QCD plasma. Semiclassical approximation}
\date{\it Institute of System Dynamics\\
and Control Theory Siberian Branch\\
of Academy of Sciences of Russia,\\
P.O. Box 1233, 664033 Irkutsk, Russia}

\thispagestyle{empty}
\maketitle{}
%%%% Making dual numbered equations %%%%%%%%%%%%

\def\theequation{\arabic{section}.\arabic{equation}}

\[
{\bf Abstract}
\]
Within the framework of the semiclassical approximation, we derive the Boltzmann equation
describing the dynamics of colorless plasmons in a hot QCD plasma.
The probability of the plasmon-plasmon scattering at the leading order in the
coupling constant is obtained. This probability
is gauge-independent at least in the class  of the covariant and temporal
gauges. It is noted that the structure of the scattering kernel possesses important
qualitative difference from the corresponding one in the Abelian plasma,
in spite of the fact that we focused our study on the colorless soft
excitations. It is shown that four-plasmon decay is suppressed by the power of
$g$ relative to the process of nonlinear scattering of plasmons by thermal
particles at the soft momentum scale. It is stated that the former
process becomes important in going to the ultrasoft region of the momentum scale.


\newpage

\section{Introduction}
\setcounter{equation}{0}

For about three decades there has been an increasing interest in theoretical
research into various dynamical properties of (ultra)relativistic many-particle
systems. It is connected with manifold applications to various problems in
astrophysical systems, modern cosmology, in multiparton processes in experiment
with high energy heavy ion collisions etc. The kinetic phenomenons,
having a purely collective character, are one of the more important
aspects of complicated dynamics of many-particle systems under extreme conditions.
Here, the basic element in the description of transport phenomena is the derivation of
the corresponding kinetic equations which would take into account (depending
on the character of the problem being studied) the presence of mean fields in the system,
two - (and more) body collisions, the possible renormalization effects,
the effects of quantum fluctuations (stochastic), pair production,
etc. Here, we restrict our consideration to a brief review of the
results of the derivation of relativistic kinetic equations essentially based on
two-body collisions in hot gauge theories.

At present there are a few methods of construction of relativistic
collision integrals. In particular, we mention
{\it the Zubarev's method of the non-equilibrium statistical operator} [1]
(use of this method on a relativistic systems of
quark-gluon plasma (QGP) type can be found in Ref. [2]), and the method developed by
Klimontovich [3] for an ordinary nonrelativistic plasma (the so-called
{\it second momentum} or {\it polarization approximation}) and expanded
to relativistic (semi)classical systems - in [4]. It should be stressed that the
above-mentioned methods are particularly effective in the construction
of collision integrals for relativistic (semi)classical systems, whose
evaluation is described by so-called exact "microscopic" dynamical equations
arising from the classical equations of motion. However, the extension of these
methods to the relativistic quantum systems encounters some difficulties and therefore
is uneffective. For the latter systems, mention may be made of the method
based on the use of the reduction formulae of quantum field theory given
by de Groot {\it et al} [5]. A more powerful and more convenient
tool to derive the approximate relativistic kinetic equations from exact field
Shwinger-Dyson equations, is the so-called {\it closed-time-path}
(CTP) formalism [6]. Examples of its relativistic
field-theoretical generalization can be found in [7, 8].

As is known, the cornerstone of derivation of the kinetic equations for
a hot non-Abelian plasma is a fundamental separation of the momentum scale.
The physical justification for such a separation is the fact that
the collective excitations which develop at a particular energy scale $g T$
(or $g^2T$, where $T$ is the temperature,
and $g$ is the coupling constant), are well separated, when $g \ll 1$, from
the typical energies of plasma particles $\sim T$ [9]. Generally
speaking one can define two types of kinetic
equations: the equations
for hard \footnote{For the connection with our previous papers [10]
here, we shall follow the definitions of the momentum scales, accepted in [18]:
the hard scale, corresponding to momentum of order $T$,
the soft scale $\sim gT$, and the ultrasoft scale $\sim g^2 T$.}
particles - hard quarks($q$), antiquarks($\bar{q}$) and hard transverse
gluons(${\rm g}$), and equation for
soft (ultrasoft) collective modes (in the case of Bose excitations - transverse and
longitudinal modes, the latter are called by plasmons). Most efforts
were directed at the derivation of the first type of equations. However for
considerably excited states, when a characteristic time of relaxation of the
hard particle distributions $f_s, s = q, \bar{q}, {\rm g}$ is commensurable with
a characteristic time of relaxation of the soft oscillations or even
significantly exceeds it, along with kinetic equations for hard particles it
is necessary to use the kinetic equation for the soft modes.

The calculation of the collision term for the quark-gluon plasma was
apparently first made in [11].
Within the framework of the concepts developed in the theory of electron-ion plasma,
the scattering probability of (anti)quarks among themselves through longitudinal
and transverse virtual oscillations which account for dynamical screening,
is derived. We note that in this paper the Boltzmann equation for hard quarks
and antiquarks was supplemented by an equation characterized the relaxation of
field excitations and was presented as the second type of kinetic equation
mentioned above. In paper [12] a similar Boltzmann equation for hard gluons
was obtained. Here, the scattering probability was deduced within the framework
of usual diagrammatic perturbation theory with inclusion of screening effects in the
random-phase approximation (one-loop order). Although the relativistic Boltzmann
equations constructed in these papers take into account such an important
QGP property as screening, the range of their validity restricts their use
to just colorless deviations from equilibrium distribution functions.

In [4] within the framework of the (semi)classical representation of QGP [13]
the Balescu-Lenard-type collision terms for small color and singlet deviations
of the distributions from the initial colorless equilibrium, were derived by
the Klimontovich method. However, in these papers, the organizing role of the various
momentum scales was not recognized, resulting in some inconsistent and complicated
transport equations for hard particles.

Only in recent years it been possible to derive the Boltzmann equation
for hard modes of hot non-Abelian plasma by a rigorous and self-consistent way
using an expansion in the
coupling constant, and clearly clarify the nature of the approximation involved,
and thus fix its range of applicability. It was be shown that for longer
wavelengths ($\lambda \sim 1/g^2 T$) of color excitations in the non-Abelian
plasma, not only was consideration of interaction of hard particles with the soft
degrees of freedom represented by mean fields essential, but also taking into
account the collisions of hard particles among themselves.
A similar Vlasov-Boltzmann equation reproduces exactly
(at leading order in $g$) a large variety of the thermal results obtained  by a
more fundamental analysis of the diagrammatic perturbation theory [14] and provides
in some cases a letter description of phenomena do not yield to a
perturbative analysis.

Here, one can extract three approaches to constructing an effective
kinetic equation for hard particles with collision term. The first of them is
connected  with B\"odeker's effective theory for the ultrarelativistic
field modes [15]. Starting from the collisionless non-Abelian Vlasov equation,
which is the result of integrating out the scale $T$ [9], B\"odeker has shown how
one can integrate out the scale $g T$ in an expansion in the gauge coupling $g$.
At leading order in $g$, he has obtained the linearized Vlasov-Boltzmann equation
for the hard field modes, which besides a collision term also contains a
Gaussian noise. Subsequently, this equation
was also proposed by Arnold {\it et al} [16] who derive the relevant
collision term on phenomenological grounds -- by analyzing the scattering
processes between hard particles in the plasma. The kinetic equation
derived in [15, 16] has a non-trivial matrix structure, since, the
distribution function that describes color fluctuations is not diagonal in
color space.

Afterwards, an alternative derivation of the collision term of Balescu-Lenard-type was
proposed by Litim and Manuel, and by Valle Basagoiti [17].
The former authors used a classical transport theory, whereas the latter used
the set of "microscopic" dynamical equations coming from the HTL effective
action describing the evolution of the collisionless plasma. In both cases
the collision terms were derived by averaging the statistical fluctuations
in the plasma on the basis of the method developed by Klimontovich [3].

Blaizot and Iancu [18] suggested a detailed derivation of the Vlasov-Boltzmann
equation, starting from the Kadanoff-Baym equations. The derivation is based
on the method of gauge covariant gradient expansion, which was first
proposed by them for the collective dynamics at the scale $g T$ [9].
By using the given equation they obtain [19] the effective amplitudes
for the ultrasoft color fields, which generalize the HTL's by including
the effects of the collisions (see also Guerin [20]).

The paper of Bezzerides and DuBois [21] devoted to the non-thermal QED plasma,
is one of the first papers, in which the relativistic kinetic equation for
the soft correlation function was considered. Over 60-70 years in
connection with application to thermonuclear fusion in the theory of the
nonrelativistic electron-ion plasma, a powerful perturbative method
(the so-called {\it weak turbulent approximation}) was developed [22-24]
for research into various
nonlinear plasma processes of the following types: on-shell scattering
of soft modes on hard particles (the nonlinear Landau damping), three- and four-wave
decays, etc.
In spite of the fact that this method is not able to describe the phenomena
connected with strong turbulence; nevertheless, it enables one within the framework of
a unified scheme to encompass a wide class of plasma phenomena. The paper [21]
is the attempt at the extending of the above-mentioned weak turbulent theory to the
electron-positron-photon plasma governed by the quantum electrodynamics. As a basic
tool for such an extension, the CTP-formalism was used. However, since the main effort here
was directed into investigating the collision integrals for hard electron and
positrons, the authors have restricted their derivation to the plasmon kinetic
equation, only taking into account pair production and the (linear) Landau
damping process. Properly high-order processes, which we are interested in and are
responsible for the nonlinear interaction mechanisms of the plasma waves,
were not considered at all (see also the discussion of one-loop computations below).

A similar kinetic equation for soft Bose-modes in a QGP was apparently first
made by Heinz {\it et al} [25]. In the context of the imaginary
time formalism in the one-loop approximation the imaginary part of the complete
color linear response function was deduced and it is shown that
it can be expressed in the form of the Boltzmann-Nordheim collision term.
On the basis of such a derived rate of decay $(\Gamma_{\rm d})$ and the
rate for regeneration of the perturbations $(\Gamma_i)$, the kinetic equation
defining the evolution of a phase space distribution $N ({\bf x}, t; {\bf k},
\omega)$ of soft electric perturbations of the momentum
$k = ( \omega, {\bf k})$ in the form proposed by Weldon [26], was written as
\begin{equation}
\frac{{\rm d} N}{{\rm d} t} = - N \Gamma_{\rm d} + (1 + N) \Gamma_i .
\label{eq:q}
\end{equation}
As in a previous case, the higher orders, when $\Gamma_{\rm d}$ and
$\Gamma_i$ itself can functionally (in the general case, nonlinearly) depend on
$N$ here, were not considered. However the derivation of such dependence
becomes important if we take into account that all computations in [25] ([21])
were performed with the "rigid" one-loop approximation, with bare propagators of
massless gluons and quarks (electrons). However, as is known, quarks (electrons)
and gluons inside the loop are not massless, they acquire the effective
temperature-induced masses. The consequence of this fact is a kinematic prohibition
of a decay of the soft perturbations into physical states. Furthermore by virtue of the
fact that the phase velocities of both transverse and longitudinal eigenmodes of
the plasma exceed velocity of light, the linear Landau damping is also absent.
By virtue of the above-mentioned, the rates of decay and regeneration are just zero
in this approximation.

This paper is devoted to further study of the kinetic equation for soft modes
of the non-Abelian plasma. The theoretical framework of this paper is derived from
synthesis of two formal developments. The first one is the development of a nonlinear
theory of plasma wave interactions in ordinary plasma - more exactly the weak turbulent approximation [22-24].
The second is the development of effective theory of hot QCD originally proposed
by Braaten and Pisarski [27], Frenkel and Taylor [28], Jackiw and Nair [29],
Blaizot and Iancu [9] and then developed in the papers [15-20].
In our previous papers [10] without resorting to a complicated diagrammatic
technique within the framework of the semiclassical representations, the following term
in the expansions of $\Gamma_{\rm d}$ and $\Gamma_i$,
linear on a phase space distribution of soft perturbations, was derived.
However, as was shown in [10] (see also section 2), this approximation is not
sufficient for a complete definition of the relaxation process of the soft
modes in QGP. Here, we consider the next terms in the expansions of
$\Gamma_{\rm d}$ and $\Gamma_i$, and show that the corresponding nonlinear
equation (\ref{eq:q}) is of purely Boltzmann type, i.e. the collision
term on the r.h.s. of this equation has a standard Boltzmann structure, with a
gain term and a loss term.

The outline of the paper is as follows. In section\,2 the preliminary comments,
with regard to derivation of the Boltzmann equation, describing the plasmon-plasmon
scattering are explained. In section\,3 the essential features of the scheme, which we
used previously in [10] to derive the kinetic equation with allowance for the
nonlinear Landau damping, are summarized. In section\,4 we discuss the consistency with gauge
symmetry of the approximation scheme used. Section\,5 is
devoted to the determination of the interacting fields in the form of the expansion in
free fields with the necessary accuracy for further research. In section\,6 we select
all terms in the expansion of the color random current responsible for the
four-plasmon decay and derive the intermediate
kinetic equation which then in section\,7 will be rewritten in the terms of HTL-amplitudes.
Section\,8 is devoted to deriving the probability of plasmon-plasmon scattering,
which is the main result of this work. In the next section on the basis of the
explicit form of the obtained collision integral, the expression for lifetimes
of colorless plasmons is defined and an estimate for the leading order in the
coupling at the soft momentum scale is deduced. Finally in section\,10 we present
our conclusions and future avenues of study.

\section{Preliminary comments}
\setcounter{equation}{0}

We denote the localized number density of the plasmons by $N^l ({\bf k}, x) \equiv
N_{\bf k}^l$, and the distribution function of hard thermal gluons by
$f ( {\bf p}, x) \equiv f_{\bf p}$. In this paper we consider processes with
longitudinal oscillations only, propagating in a purely gluonic plasma, with
no quarks. Besides, we suppose that there is no external color current
and/or mean color field in the system, and the system is in the global
equilibrium state, i.e.
\begin{equation}
f_{\bf p}^{ab} = \delta^{ab} f_{\bf p} \equiv \delta^{ab} \, 2 \,
\frac{1}{{\rm e}^{E_{\bf p}/T} - 1} \, .
\label{eq:q1}
\end{equation}
Here, $E_{\bf p} \equiv \vert {\bf p} \vert$ for a massless hard gluon,
the coefficient 2 takes into account that the hard gluon has two
helicity states and $a,b = 1, \ldots , N_c^2 - 1$ for the $SU(N_c)$ gauge group.
The triviality of the color structure of the plasmon number
density is a consequence of these restrictions (see section 4)
\[
N_{\bf k}^{l \, ab} = \delta^{ab} N_{\bf k}^l .
\]
The dispersion relation for plasmons $\omega = \omega^l ({\bf k}) \equiv
\omega_{\bf k}^l$ is defined from
\begin{equation}
{\rm Re} \, \varepsilon^l (\omega, {\bf k}) = 0 ,
\label{eq:w1}
\end{equation}
where
\begin{equation}
\varepsilon^l (\omega, {\bf k}) = 1 + \frac{3 \omega_{pl}^2}{{\bf k}^2}
\bigg[1 - F \bigg( \frac{\omega}{\vert {\bf k} \vert} \bigg) \bigg] \; , \;
F (x) \equiv \frac{x}{2} \bigg[ \ln \bigg \vert \frac{1 + x}{1 - x}
\bigg \vert - i \pi
\theta ( 1 - \vert x \vert ) \bigg]
\label{eq:e1}
\end{equation}
is longitudinal color permeability and
$\omega_{pl}^2 = g^2 N_cT^2/9$ is a plasma frequency.

We expect the time-space evolution of $N_{\bf k}^l$ to be described by
\begin{equation}
\frac{\partial N_{\bf k}^l}{\partial t} +
{\bf V}_{\bf k}^l \frac{\partial N_{\bf k}^l}{\partial {\bf x}} =
- N_{\bf k}^l \Gamma_{\rm d} [ N_{\bf k}^l ] + ( 1 + N_{\bf k}^l )
\Gamma_i [ N_{\bf k}^l ] ,
\label{eq:r1}
\end{equation}
where ${\bf V}_{\bf k}^l = \partial \omega_{\bf k}^l/\partial {\bf k}$ is
the group velocity of the longitudinal oscillations\footnote{Notice, that in the
general case, when the distribution function of hard gluons is a
slowly varying function in time and space, equation (\ref{eq:w1}) is replaced
by ${\rm Re} \, \varepsilon^l (\omega, {\bf k}; t, {\bf x}) = 0$. In this case
the l.h.s. of equation (\ref{eq:r1}) should be supplemented by
$( \partial \omega^{l}({\bf k},x) / \partial {\bf x})
(\partial N^l({\bf k},x) / \partial {\bf k})$.}. For a generalized decay rate
$\Gamma_{\rm d}$ and inverse decay rate $\Gamma_i$ it is shown in an explicit
form that in general case they are functionals dependent on the plasmon number
density. Although the approach we shall use in the subsequent discussion
is correct only within the framework of
semiclassical approximation, it is convenient to interpret the terms
entering into $\Gamma_{\rm d}$ and $\Gamma_i$, using a quantum language.

The Eq.\,(\ref{eq:r1}) in general, describes two principal
processes of the nonlinear wave-interaction. The first of them represents the process of the stimulated
emission and absorption of the collective wave quanta by hard particles of
plasma. In this case the more general expression for the decay rate
$\Gamma_{\rm d}^{({\cal S})}$ and the regeneration rate $\Gamma_i^{({\cal S})}$
can be written in the following forms, respectively:
\begin{equation}
\Gamma_{\rm d}^{({\cal S})} [N_{\bf k}^l] = \sum_{n, m} \int
\frac{{\rm d} {\bf p}}
{(2 \pi)^3} \int {\rm d} {\cal T}_{nm}^{({\cal S})} \, {\it w}
({\bf p} \vert {\bf k}, {\bf k}_1, \ldots ,{\bf k}_n;
{\bf k}_1^{\prime}, \ldots , {\bf k}_m^{\prime})
\, N_{{\bf k}_1}^l \ldots N_{{\bf k}_n}^l
\label{eq:t1}
\end{equation}
\[
\times (1 + N_{{\bf k}_1^{\prime}}^l) \ldots
(1 + N_{{\bf k}_m^{\prime}}^l)
f_{\bf p} [1 + f_{{\bf p}^{\prime}}] ,
\]
and
\begin{equation}
\Gamma_i^{({\cal S})} [N_{\bf k}^l] = \sum_{n, m} \int  \frac{{\rm d} {\bf p}}
{(2 \pi)^3} \int {\rm d} {\cal T}_{nm}^{({\cal S})} \, {\it w}
({\bf p}^{\prime} \vert {\bf k}_1^{\prime}, \ldots ,{\bf k}_m^{\prime};
{\bf k}, {\bf k}_1, \ldots , {\bf k}_n)
\, N_{{\bf k}_1^{\prime}}^l \ldots N_{{\bf k}_m^{\prime}}^l
\label{eq:y1}
\end{equation}
\[
\times (1 + N_{{\bf k}_1}^l) \ldots (1 + N_{{\bf k}_n}^l)
f_{{\bf p}^{\prime}} [1 + f_{\bf p}] .
\]
Here, the phase-space integration is
\begin{equation}
\int {\rm d} {\cal T}_{nm}^{({\cal S})} \equiv
\int \frac{{\rm d} {\bf k}_1}{(2 \pi)^3} \ldots
\frac{{\rm d} {\bf k}_n}{(2 \pi)^3}
\frac{{\rm d} {\bf k}_1^{\prime}}{(2 \pi)^3} \ldots
\frac{{\rm d} {\bf k}_m^{\prime}}{(2 \pi)^3}
\label{eq:u1}
\end{equation}
\[
\times (2 \pi) \, \delta (E_{\bf p} + \omega_{\bf k}^l + \omega_{{\bf k}_1}^l + \ldots +
\omega_{{\bf k}_n}^l -
E_{{\bf p}^{\prime}} - \omega_{{\bf k}_1^{\prime}}^l - \ldots -
\omega_{{\bf k}_m^{\prime}}^l) ,
\]
with the delta function expressing the energy conservation of the processes
of stimulated emission and absorption of the plasmons. The function
${\it w}({\bf p} \vert {\bf k}, {\bf k}_1, \ldots, {\bf k}_n;
{\bf k}_1^{\prime},
\ldots , {\bf k}_m^{\prime})$ is the probability of absorption of $n + 1$ plasmons
with the frequencies
$\omega_{\bf k}^l, \omega_{{\bf k}_1}^l, \ldots , \omega_{{\bf k}_n}^l$
and the wavevectors ${\bf k}, {\bf k}_1, \ldots, {\bf k}_n$
by a hard gluon carrying of momentum ${\bf p}$ with consequent
radiation of $m$ plasmons with frequencies
$\omega_{{\bf k}_1^{\prime}}^l, \ldots, \omega_{{\bf k}_m^{\prime}}^l$
and the wavevectors ${\bf k}_1^{\prime}, \ldots, {\bf k}_m^{\prime}$.
The function  ${\it w} ( {\bf p}^{\prime} \vert {\bf k}_1^{\prime}, \ldots ,
{\bf k}_m^{\prime}; {\bf k}, {\bf k}_1, \ldots , {\bf k}_n)$ is the probability of
inverse process - the absorption of $m$ plasmons by a hard gluon with the momentum
${\bf p}^{\prime} \equiv {\bf p} +  {\bf k} + {\bf k}_1 + \ldots +
{\bf k}_n - {\bf k}_1^{\prime} - \ldots - {\bf k}_m^{\prime}$ with consequent
radiation of $n + 1$ plasmons. Diagrammatically this corresponds to a Feynman graph
with two hard external lines and an arbitrary number of $n + m +1$ soft
external lines. By using the fact that $\vert {\bf p} \vert
\gg \vert {\bf k} \vert,
\vert {\bf k}_1 \vert, \ldots , \vert {\bf k}_n \vert ,
\vert{\bf k}_1^{\prime} \vert,
\ldots , \vert {\bf k}_m^{\prime} \vert$, the energy conservation law can be
represented in the form of following "generalized" resonance condition:
\begin{equation}
\omega_{\bf k}^l + \omega_{{\bf k}_1}^l + \ldots + \omega_{{\bf k}_n}^l -
\omega_{{\bf k}_1^{\prime}}^l - \ldots - \omega_{{\bf k}_m^{\prime}}^l -
{\bf v} ({\bf k} + {\bf k}_1 + \ldots + {\bf k}_n - {\bf k}_1^{\prime} -
\ldots - {\bf k}_m^{\prime}) = 0,
\label{eq:i1}
\end{equation}
where ${\bf v} \equiv {\bf p}/\vert {\bf p} \vert$ .
Furthermore, one can approximate the distribution function of hard gluons
on the r.h.s. of equations (\ref{eq:t1}) and (\ref{eq:y1}),
\[
f_{{\bf p}^{\prime}} \simeq f_{\bf p} + ( {\bf k} + {\bf k}_1 + \ldots +
{\bf k}_n - {\bf k}_1^{\prime} - \ldots
-{\bf k}_m^{\prime}) \frac{\partial f_{\bf p}}{\partial {\bf p}} ,
\]
and set $1 + f_{\bf p} \simeq 1 + f_{{\bf p}^{\prime}} \simeq 1$ by virtue  of
$f_{\bf p}, f_{{\bf p}^{\prime}} \ll 1$.

The r.h.s. of (\ref{eq:t1}) and (\ref{eq:y1}) can be formally considered as
an expansions of $\Gamma_{\rm d}^{({\cal S})}$ and $\Gamma_i^{({\cal S})}$
in the functional
series in powers of the plasmon number density. The actual dimensionless
parameter of expansion here, is (for classical statistic) the ratio of the energy of
longitudinal plasma excitations to the averaged thermal energy per particle,
i.e.
\[
\varepsilon = \bigg( \int \frac{{\rm d} \, {\bf k}}{(2 \pi)^3} \,
\omega_{\bf k}^l
N_{\bf k}^l \bigg)  \biggm/ \bigg( \bar{n} \int \frac{{\rm d} \,
{\bf p}}{(2 \pi)^3} \, E_{\bf p}
f_{\bf p} \bigg),
\]
where $\bar{n}$ is the mean density. In conditions, when the excitations energy
is a small quantity compared with the thermal energy of hard particles, we have
\begin{equation}
\varepsilon \ll 1 .
\label{eq:o1}
\end{equation}
The last inequality means that the fields of longitudinal oscillations
are sufficiently small and they cannot essentially change such "crude" equilibrium
parameters of a plasma as particles density, temperature and thermal energy
(this, in particular, justifies the choice of the distribution function of
thermal gluons in the form of (\ref{eq:q1})).

On the other hand, however, we shall consider the energy of the plasma
oscillations to be sufficiently large, i.e. greatly exceeding the energy of thermal fluctuations of
the the color field in the plasma. The consequence of the last requirement is the
inequality
\begin{equation}
\varepsilon \gg \delta ,
\label{eq:p1}
\end{equation}
where $\delta$ is the plasma parameter
\[
\delta = \frac{\bar{r}^3}{r_D^3} \ll 1 .
\]
Here, $\bar{r}$ is the inter-particle distance $(\sim \bar{n}^{-1/3})$, and
$r_D$ is the Debye length
\[
r_D^2 = \frac{T}{4 \pi \bar{n} g^2 N_c} \, .
\]
The condition (\ref{eq:p1}) demonstrates the validity of ignoring the hard gluon
collisions among themselves relative to their interactions with soft plasma
modes.

Inequalities (\ref{eq:o1}) and (\ref{eq:p1}) correspond to the weak turbulent
approximation, within the framework of which one can restrict the consideration
to several first terms in a functional expansions of
$\Gamma_{\rm d}^{({\cal S})}$ and $\Gamma_i^{({\cal S})}$.
We note however, that when the energy level of the plasma excitations becomes
comparable with the thermal energy of particles, e.g. as the result of development of
a strong instability ({\it strong turbulence}), the perturbation
theory here, is no longer applicable and the problem of summation of all
the relevant contributions thus appears.
The last situation can be really take place in the processes proceed in QGP
emerging from the heavy ion collisions at higher energies. In this work
we do not consider this very complicated problem, assuming that the inequality
(\ref{eq:o1}) is always fulfilled.

Let us discuss in more detail the first terms in the expansions of
$\Gamma_{\rm d}^{({\cal S})}$ and $\Gamma_i^{({\cal S})}$. For $n = m = 0$, the equation (\ref{eq:i1}) results in
the relation
\[
\omega_{\bf k}^l - {\bf v}{\bf k} = 0 ,
\]
which is well-known as {\it the Cherenkov resonance condition}, which does not hold in the gluon plasma.
Therefore $\Gamma_{\alpha}^{({\cal S})} = \Gamma_i^{({\cal S})} = 0$, when
only $O (\varepsilon^0)$ terms on the r.h.s. of (\ref{eq:t1}) and (\ref{eq:y1})
are retained.

For $n = 1, m = 0$ we have
\begin{equation}
\omega^l_{\bf k} + \omega^l_{{\bf k}_1} - {\bf v}({\bf k} +{\bf k}_1)=0,
\label{eq:a1}
\end{equation}
and for $n = 0, m = 1$, respectively (here, we replace ${\bf k}_1^{\prime}$ by
${\bf k}_1$)
\begin{equation}
\omega^l_{\bf k} - \omega^l_{{\bf k}_1} - {\bf v}({\bf k} -{\bf k}_1)=0.
\label{eq:s1}
\end{equation}
The first resonance condition (\ref{eq:a1}) describes the simultaneous radiation
(or absorption) of two plasmons with frequencies $\omega_{\bf k}^l, \,
\omega_{{\bf k}_1}^l$ and wavevectors ${\bf k}, \, {\bf k}_1$. By virtue of the fact
that the phase velocity of the longitudinal oscillations exceeds the velocity of light,
this process is kinematically forbidden. The first non-trivial terms in the
expansion of the functionals $\Gamma_{\rm d}^{({\cal S})}$ and
$\Gamma_i^{({\cal S})}$ are defined
by the second resonance condition (\ref{eq:s1}). It is associated with the absorption
of the plasmon by a hard gluon with frequency $\omega_{\bf k}^l$ and
wavevector ${\bf k}$ with its consequent radiation with frequency
$\omega_{{\bf k}_1}^l$ and wavevector ${\bf k}_1$ (and vice versa).
Schematically this process can be represented as follows:
\begin{equation}
{\rm g}^{\ast} + {\rm g} \rightleftharpoons  {\rm g}^{\ast} + {\rm g} ,
\label{eq:d1}
\end{equation}
where ${\rm g}^{\ast}$ are the plasmon collective excitations and ${\rm g}$ are excitations
with characteristic momenta of order $T$. In the theory of the ordinary plasma
[22-24] this process is known as {\it the nonlinear Landau damping}. In the case of
QGP it was studied in detail in [10]. We have shown, that the nonlinear Landau
damping rate
\begin{equation}
\gamma^l ({\bf k}) \equiv ( \Gamma_{\rm d}^{({\cal S})} [ N_{\bf k}^l ] -
\Gamma_i^{({\cal S})} [ N_{\bf k}^l ]) \vert_{n = 0, \, m = 1}
\label{eq:f1}
\end{equation}
defines two processes: the effective pumping of energy across the spectrum
towards small wavenumbers with the conservation of excitation
energy and properly nonlinear dissipation (damping) of the longitudinal plasma
waves by hard particles, where the first process is crucial. The consequence of
this fact is the inequality: $\gamma^l (0) < 0$, i.e. ${\bf k} = 0$ - mode is
increased. The main conclusion, which we drew in [10] is that the only process
of the nonlinear Landau damping does not lead to the total relaxation of soft
excitations in the gomogeneous isotropic plasma. At the scale of a small
$\vert {\bf k} \vert \; (\vert {\bf k} \vert \ll g T)$ it is necessary
to consider the processes of higher-order in $\varepsilon$, than ({\ref{eq:d1}), which
lead to the suppression of increase of the ${\bf k} = 0$-mode.

The following terms in the expansion of the decay rate (\ref{eq:t1}) and the
inverse decay rate (\ref{eq:y1}), corresponding to $n, m = 1, 2$, are defined by
\begin{equation}
\omega^l_{\bf k} \pm \omega^l_{{\bf k}_1} \pm \omega^l_{{\bf k}_2}
- {\bf v}({\bf k} \pm {\bf k}_1 \pm {\bf k}_2)=0,
\label{eq:g1}
\end{equation}
\[
\omega^l_{\bf k} \pm \omega^l_{{\bf k}_1} \mp \omega^l_{{\bf k}_2}
- {\bf v}({\bf k} \pm {\bf k}_1 \mp {\bf k}_2)=0.
\]
Physically this corresponds to simultaneous absorption (radiation) of three
plasmons by a thermal gluon, or simultaneous absorption (radiation) of two plasmons
with consequent radiation (absorption) of one plasmon. At the long-wavelength
range these processes are kinematically forbidden and therefore in this approximation
$\Gamma_{\rm d}^{({\cal S})}$ and $\Gamma_i^{({\cal S})}$ vanish.

However there are contributions different from zero in the expansions
of the generalized decay rate $\Gamma_{\rm d}$ and the inverse
$\Gamma_i$ in the equation (\ref{eq:r1}) which are of the same order as the
processes (\ref{eq:g1}), i.e. of order
$O (\varepsilon^2)$. These contributions
are concerned with the second type of the nonlinear processes defining the
time-space evolution of $N_{\bf k}^l$ and going without exchange of energy
between hard thermal gluons and plasmons. They represent the processes
of decays, fusions of plasmons and their scattering off each other. Diagrammatically,
this corresponds to Feynman graph, where all external lines are soft.
The relevant decay rate $\Gamma_i^{({\cal P})}$ and regenerating rate
$\Gamma_{\rm d}^{({\cal P})}$ can be formally represented in the form
\begin{equation}
\Gamma_{\rm d}^{({\cal P})} [N_{\bf k}^l] = \sum_{n, m}
\int {\rm d} {\cal T}_{nm}^{({\cal P})} \, {\it w}
({\bf k}, {\bf k}_1, \ldots ,{\bf k}_n;
{\bf k}_1^{\prime}, \ldots , {\bf k}_m^{\prime})
\, N_{{\bf k}_1}^l \ldots N_{{\bf k}_n}^l
(1 + N_{{\bf k}_1^{\prime}}^l) \ldots
(1 + N_{{\bf k}_m^{\prime}}^l) ,
\label{eq:h1}
\end{equation}
and
\begin{equation}
\Gamma_i^{({\cal P})} [N_{\bf k}^l] = \sum_{n, m}
\int {\rm d} {\cal T}_{nm}^{({\cal P})} \, {\it w}
({\bf k}_1^{\prime}, \ldots ,{\bf k}_m^{\prime};
{\bf k}, {\bf k}_1, \ldots , {\bf k}_n)
\, N_{{\bf k}_1^{\prime}}^l \ldots N_{{\bf k}_m^{\prime}}^l
(1 + N_{{\bf k}_1}^l) \ldots (1 + N_{{\bf k}_n}^l).
\label{eq:j1}
\end{equation}
Here, the phase-space measure is
\begin{equation}
\int {\rm d} {\cal T}_{nm}^{({\cal P})} \equiv
\int \frac{{\rm d} {\bf k}_1}{(2 \pi)^3} \ldots
\frac{{\rm d} {\bf k}_n}{(2 \pi)^3}
\frac{{\rm d} {\bf k}_1^{\prime}}{(2 \pi)^3} \ldots
\frac{{\rm d} {\bf k}_m^{\prime}}{(2 \pi)^3}
\, (2 \pi)^4 \delta^{(4)} (k + k_1 + \ldots + k_n - k_1^{\prime} -
\ldots - k_m^{\prime}),
\label{eq:k1}
\end{equation}
with the delta functions expressing the energy and the momentum conservation of
the decay processes, the fusion and the plasmon scattering among themselves.
The decay rate
(\ref{eq:h1}) and the regenerating rate (\ref{eq:j1}) are different
from zero if the following "resonance" conditions are obeyed
\[
\omega_{\bf k}^l + \omega_{{\bf k}_1}^l + \ldots + \omega_{{\bf k}_n}^l -
\omega_{{\bf k}_1^{\prime}}^l - \ldots - \omega_{{\bf k}_m^{\prime}}^l = 0,
\]
\[
{\bf k} + {\bf k}_1 + \ldots + {\bf k}_n - {\bf k}_1^{\prime} -
\ldots - {\bf k}_m^{\prime} = 0.
\]
The first contribution is different  from zero in $\Gamma_{\rm d}^{({\cal P})}$
and $\Gamma_i^{({\cal P})}$, which would arise in the case of
$n = m = 1, 2$ as defined by the system of equations
\begin{equation}
\omega_{\bf k}^l \pm \omega_{{\bf k}_1}^l \mp \omega_{{\bf k}_2}^l = 0,
\label{eq:l1}
\end{equation}
\[
{\bf k} \pm {\bf k}_1 \mp {\bf k}_2 = 0.
\]
These conservation laws describe a decay of one plasmon into two plasmons
and the reverse process of fusion of two plasmons into one plasmon,
\begin{equation}
{\rm g}^{\ast} \rightleftharpoons  {\rm g}^{\ast}_1 + {\rm g}^{\ast}_2.
\label{eq:z1}
\end{equation}
Three-plasmon decay (\ref{eq:z1}) is as important as the process of nonlinear
Landau damping (\ref{eq:d1}). However, the specific peculiarity of
a dispersion law of the longitudinal oscillations in hot non-Abelian plasma
is that resonance equations (\ref{eq:l1}) have no solutions no
matter what the values of wavevectors ${\bf k}, {\bf k}_1$
and ${\bf k}_2$ may be.
The processes (\ref{eq:z1}) are kinematically forbidden by the conservation laws,
and it makes a contribution only in the second order over parameter
$\varepsilon$ of perturbation theory, for $n, m = 1, 2, 3$.

In general case, with four plasmons, two different processes occur
\begin{equation}
{\rm g}^{\ast} \rightleftharpoons {\rm g}_1^{\ast} + {\rm g}_2^{\ast} +
{\rm g}_3^{\ast} ,
\label{eq:x1}
\end{equation}
\begin{equation}
{\rm g}^{\ast} + {\rm g}_1^{\ast}  \rightleftharpoons {\rm g}_2^{\ast} +
{\rm g}_3^{\ast} .
\label{eq:c1}
\end{equation}
The first of them corresponds to the process of the decay of one plasmon
${\rm g}^{\ast}$
into three plasmons ${\rm g}_1^{\ast}, \, {\rm g}_2^{\ast}, \, {\rm g}_3^{\ast}$,
and the reverse process of fusion of three plasmons into one plasmon ${\rm g}^{\ast}$.
The second process presents the plasmon scattering by plasmon. The last one
is considered as the decay process and is interpretated as the process of the decay
(fusion) of two plasmons
${\rm g}^{\ast}$ and ${\rm g}_1^{\ast}$ into two plasmons ${\rm g}_2^{\ast}$ and
${\rm g}_3^{\ast}$.
For four-plasmon decay process (\ref{eq:x1}), the following
resonance conditions are obeyed:
\begin{equation}
\omega_{\bf k}^l - \omega_{{\bf k}_1}^l - \omega_{{\bf k}_2}^l -
\omega_{{\bf k}_3}^l = 0,
\label{eq:v1}
\end{equation}
\[
{\bf k} - {\bf k}_1 - {\bf k}_2 - {\bf k}_3 = 0,
\]
and for the scattering process (\ref{eq:c1}) we have, respectively,
\begin{equation}
\omega_{\bf k}^l + \omega_{{\bf k}_1}^l - \omega_{{\bf k}_2}^l -
\omega_{{\bf k}_3}^l = 0,
\label{eq:b1}
\end{equation}
\[
{\bf k} + {\bf k}_1 - {\bf k}_2 - {\bf k}_3 = 0.
\]
It is not difficult to show that the conservation laws (\ref{eq:v1}) and
(\ref{eq:b1}) kinematically
forbid the processes of the direct decay of the plasmon into three (and
vice versa) (\ref{eq:x1}) and admit only the processes of (\ref{eq:c1}) type.
As shown in section 9, at the soft scale the latter process is suppressed
by a power of $g$ relative to the process of nonlinear Landau damping
(\ref{eq:d1}). However at the ultrasoft scale, one would expect that the
process of the elastic scattering of the plasmon by a plasmon may be as larger as
the process (\ref{eq:d1}) and thus plays an important role
in the kinetics of the plasmons at larger wavelength (see the end of section 9).

Putting the expressions (\ref{eq:h1}) and (\ref{eq:j1}) into equation (\ref{eq:r1}),
where only $O (\varepsilon^2)$ relevant terms in the expansions are retained,
we result in the Boltzmann equation, describing four-plasmon
decays of (\ref{eq:c1}) type
\begin{equation}
\frac{\partial N_{\bf k}^{l}}{\partial t} +
{\bf V}_{\bf k}^{l} \, \frac{\partial N_{\bf k}^{l}} {\partial {\bf x}} =
\label{eq:n1}
\end{equation}
\[
= \int \frac{{\rm d} {\bf k}_1}{(2 \pi)^3} \frac{{\rm d} {\bf k}_2}{(2 \pi)^3}
\frac{{\rm d} {\bf k}_3}{(2 \pi)^3} \,
(2 \pi)^4 \delta (\omega_{\bf k}^l + \omega_{{\bf k}_1}^l - \omega_{{\bf k}_2}^l -
\omega_{{\bf k}_3}^l)
\delta ({\bf k} + {\bf k}_1 - {\bf k}_2 - {\bf k}_3)
\]
\[
\times w ( {\bf k}, {\bf k}_1; {\bf k}_2, {\bf k}_3 )
\{ N_{{\bf k}_2}^l N_{{\bf k}_3}^l ( 1+ N_{\bf k}^l )
( 1 + N_{{\bf k}_1}^l ) - N_{\bf k}^l N_{{\bf k}_1}^l
( 1 + N_{{\bf k}_2}^l ) ( 1 + N_{{\bf k}_3}^l ) \}
\]
\[
\approx \int \frac{{\rm d} {\bf k}_1}{(2 \pi)^3} \frac{{\rm d} {\bf k}_2}{(2 \pi)^3}
\frac{{\rm d} {\bf k}_3}{(2 \pi)^3} \,
(2 \pi)^4 \delta (\omega_{\bf k}^l + \omega_{{\bf k}_1}^l - \omega_{{\bf k}_2}^l -
\omega_{{\bf k}_3}^l)
\delta ({\bf k} + {\bf k}_1 - {\bf k}_2 - {\bf k}_3)
\]
\[
\times w ( {\bf k}, {\bf k}_1; {\bf k}_2, {\bf k}_3 )
\{ N_{{\bf k}_1}^l N_{{\bf k}_2}^l N_{{\bf k}_3}^l +
N_{\bf k}^l N_{{\bf k}_2}^l
N_{{\bf k}_3}^l - N_{\bf k}^l N_{{\bf k}_1}^l N_{{\bf k}_2}^l -
N_{\bf k}^l N_{{\bf k}_1}^l N_{{\bf k}_3}^l \} .
\]
In writing this equation, we have used the fact that the probabilities
of direct and reverse
processes are equal and besides in the last line in the semiclassical
regime we consider the soft modes to be strongly populated, i.e.
$(1 + N_{\bf k}^l)(1 + N_{{\bf k}_1}^l) \approx N_{\bf k}^l N_{{\bf k}_1}^l
+ N_{\bf k}^l + N_{{\bf k}_1}^l$ etc..
A similar Boltzmann equation for plasmons was studied intensively
for the ordinary plasma [30, 31] (in [24, 30] the explicit expression for the
function ${\it w} ({\bf k}, {\bf k}_1; {\bf k}_2, {\bf k}_3)$ can be found).
The main purpose of this work is the derivation in the explicit form
of the probability of plasmon-plasmon scattering for hot non-Abelian plasma.
The function of $w ({\bf k}, {\bf k}_1; {\bf k}_2, {\bf k}_3)$
must satisfy the symmetry relations over permutation of arguments
\begin{equation}
w ({\bf k}, {\bf k}_1; {\bf k}_2, {\bf k}_3) =
w ({\bf k}_2, {\bf k}_3; {\bf k}_, {\bf k}_1) =
w ({\bf k}, {\bf k}_1; {\bf k}_3, {\bf k}_2) =
w ({\bf k}_1, {\bf k}; {\bf k}_2, {\bf k}_3) ,
\label{eq:m1}
\end{equation}
which are the consequence of the indistinguishable of the plasmons (recall that here,
we discuss only colorless plasmons, i.e. those for wich the occupation number does not
carry adjoint color indices).

In conclusion of this section we note some general properties of the processes
of four-plasmon decays. Multiplying the r.h.s. of Eq. (\ref{eq:n1})
in turn by $\omega_{\bf k}^l$ and ${\bf k}$, integrating with respect to the
wavevector ${\bf k}$, and taking into account (\ref{eq:m1}), it is easily
checked that
\[
{\cal E} \equiv \int \frac{{\rm d}{\bf k}}{(2 \pi)^3} \;
\omega^l_{\bf k} N^l_{\bf k} =  const, \; \;
{\bf {\cal K}}
\equiv \int \frac{{\rm d}{\bf k}}{(2 \pi)^3} \;
{\bf k} \, N^l_{\bf k} =  const.
\]
These relations are evident a consequence of the conservation laws of
energy and momentum in the plasmon-plasmon scattering.

On the other hand, the total plasmon numbers in such decays
should also be conserved, because the only process in which two plasmons decay
into two others is permitted.
Really, by integrating  (\ref{eq:n1}) over all ${\bf k}$-space and
taking into account the relations (\ref{eq:m1}), it is easy to verify that
\[
{\cal N} \equiv \int \frac{{\rm d}{\bf k}}{(2 \pi)^3} \;
N^l_{\bf k} = const,
\]
i.e. four-plasmon decay not changes the total number of plasmons.

\section{\bf The random-phase approximation}
\setcounter{equation}{0}

In this section let us briefly recall the main methodological notion
used in study of the processes of nonlinear wave-interaction in a non-Abelian plasma
within semiclassical approximation in [10].

We use the metric $g^{\mu \nu} = diag(1,-1,-1,-1)$ and choose units such
that $c=k_{B}=1$. The gauge field potentials are $N_{c} \times N_{c}$-matrices
in a color space defined by $A_{\mu}=A_{\mu}^{a}t^{a}$ with $N_{c}^{2}-1$
Hermitian generators of the $SU(N_{c})$ group in the fundamental representation.
The field strength tensor $F_{\mu \nu}=F_{\mu \nu}^{a}t^{a}$ with
\begin{equation}
F_{\mu \nu}^{a} = \partial_\mu A_{\nu}^{a} - \partial_\nu A_{\mu}^{a}+
gf^{abc}A_{\mu}^{b}A_{\nu}^{c}
\label{eq:q2}
\end{equation}
obeys the Yang-Mills (YM) equation in a covariant gauge
\begin{equation}
\partial_\mu F^{\mu \nu}(X) - ig[A_{\mu}(X),F^{\mu \nu}(X)] -
\xi^{-1} \partial^\nu \partial^\mu A_{\mu}(X) = -j^{\nu}(X),
\label{eq:w2}
\end{equation}
where $\xi$ is a gauge parameter. $j^{\nu}$ is a color current
\begin{equation}
j^{\nu} = gt^{a} \int d^{4}p \, p^{\nu}{\rm Tr} \, (T^{a}f),
\label{eq:e2}
\end{equation}
where $T^{a}$ are Hermitian generators of $SU(N_{c})$ group in the
adjoint representation
$((T^{a})^{bc}=-if^{abc}, {\rm Tr}(T^{a}T^{b})=N_{c} \delta^{ab})$.
The distribution function of gluons $f$ satisfies the dynamical
equation which in the semiclassical limit (when polarization effects
are neglected [32]) is [13]
\begin{equation}
p^{\mu} \tilde{\cal D}_{\mu}f
+ \frac{1}{2}gp^{\mu} \{ {\cal F}_{\mu
\nu}, \frac{\partial f}{\partial p_{\nu}} \} = 0,
\label{eq:r2}
\end{equation}
where $\tilde{\cal D}_{\mu}$ is a covariant
derivative acting as
\[
 \tilde{\cal D}_{\mu} = \partial_{\mu} - ig[{\cal A}_{\mu}(X), \; \cdot \; ],
\]
with [ , ] and $ \{ , \} $ denoting the commutator and anticommutator, respectively, and
${\cal A}_{\mu}$, ${\cal F}_{\mu \nu}$ are defined as ${\cal A}_{\mu}=
A_{\mu}^{a}T^{a}, {\cal F}_{\mu \nu} = F_{\mu \nu}^{a}T^{a}$.

The distribution function $f$
can be decomposed into two parts: regular and random, where the latter is
generated by spontaneous fluctuations in the plasma
\begin{equation}
f = f^{R} + f^{T},
\label{eq:rr2}
\end{equation}
so that
\[
\langle f \rangle = f^{R} \;, \; \langle f^{T} \rangle = 0.
\]
Here, angular brackets $ \langle \cdot \rangle$ indicate a statistical
ensemble
of averaging. The initial values of parameters
which characterize the collective degree of plasma freedom is
a such statistical ensemble. For
almost linear collective motion to be considered below this may be
initial values of oscillation phases.

We also use the definition
\begin{equation}
A_{\mu} = A_{\mu}^{R} + A_{\mu}^{T} \;,\; \langle A_{\mu}^{T} \rangle = 0.
\label{eq:rrr2}
\end{equation}
The regular (background) part of the field $A_{\mu}^{R}$
will be considered to be equal to zero. The condition for which the last
assumption holds, will be considered closely in the next section.

By averaging equation (\ref{eq:r2}) over statistical ensemble,
we obtain the kinetic equation for the regular part of the
distribution function of hard gluons $f^{R}$
\[
p^{\mu} \partial_{\mu} f^{R} = igp^{\mu}
\langle [{\cal A}_{\mu}^{T},f^{T}] \rangle - \frac{1}{2}gp^{\mu}
\langle \{ ({\cal F}_{\mu \nu}^{T})_{L}, \frac{\partial f^{T}}
{\partial p_{\nu}} \} \rangle
- \frac{1}{2}gp^{\mu}
\{ \langle ({\cal F}_{\mu \nu}^{T})_{NL} \rangle, \frac{\partial f^{R}}
{\partial p_{\nu}} \} -
\]
\begin{equation}
- \frac{1}{2}gp^{\mu}
\langle \{ ({\cal F}_{\mu \nu}^{T})_{NL}, \frac{\partial f^{T}}
{\partial p_{\nu}} \} \rangle .
\label{eq:tt2}
\end{equation}
Here, the indices $L$ and $NL$ denote the linear and nonlinear parts
with respect to field $A_{\mu}^{a}$ of the strength tensor (\ref{eq:q2}).
The correlation functions on the r.h.s. of this equation are
collision terms due to particle-wave interactions and describe the
backreaction of the background state from the plasma waves.

We assume that the typical time the nonlinear relaxation for
the oscillations is a small quantity relative to the time scale over which
the distribution of hard transverse gluons $f^{R}$ vary substantially.
Therefore we neglect by change of the regular part of the distribution
function with space and time, assuming that this function is specified and
describes the global equilibrium in the gluon plasma
\begin{equation}
f^{R} \equiv f^{0} = 2
\frac{2 \theta(p_{0})}{(2 \pi)^{3}} \delta(p^{2})
\frac{1}{{\rm e}^{(pu)/T } - 1},
\label{eq:t2}
\end{equation}
where $u_{\mu}$ is the 4-velocity of the plasma. (Here, for convenience,
we somewhat overdetermine the equilibrium
distribution function of thermal gluons (\ref{eq:q1}).)

We use the expansion in powers of the oscillations amplitude of the random function
$f^{T}$ to investigate non-equilibrium processes in QGP,
such that the excitation energy of waves is small quantity in
relation to the total energy of the particles
\begin{equation}
f^{T}= \sum_{n=1}^{\infty} f^{T(n)},
\label{eq:y2}
\end{equation}
where $f^{T(n)}$ collects the contributions of the $n$-th power in $A_{\mu}^{T}$.
The expansion of a color current, corresponding to (\ref{eq:y2}) has the form
\begin{equation}
j_{\mu}=j_{\mu}^{R} + j_{\mu}^{T} \;, \; \langle j_{\mu} \rangle =
j_{\mu}^{R} \;, \; j_{\mu}^{T}= \sum_{n=1}^{\infty} j_{\mu}^{T(n)},
\label{eq:u2}
\end{equation}
where by the definition (\ref{eq:e2}), we have
\begin{equation}
j_{\mu}^{T(n)}= gt^{a} \int d^{4}p \, p_{\mu}{\rm Tr} \, (T^{a}f^{T(n)}).
\label{eq:i2}
\end{equation}
The regular part of a current vanishes for the global equilibrium gluon plasma.

Substituting the expansion (\ref{eq:y2}) into (\ref{eq:r2}), and collecting
the terms of the same order in $A^T_{\mu}$, we derive the
system of equations
\begin{equation}
p^{\mu} \partial_{\mu} f^{T(1)} = - \frac{1}{2}gp^{\mu}
\{ ({\cal F}_{\mu \nu}^{T})_{L},
\frac{\partial f^{R}}
{\partial p_{\nu}} \},
\label{eq:o2}
\end{equation}
\begin{equation}
p^{\mu} \partial_{\mu} f^{T(2)} = igp^{\mu}(
[{\cal A}_{\mu}^{T},f^{T(1)}] -
\langle [{\cal A}_{\mu}^{T},f^{T(1)}] \rangle)
\label{eq:p2}
\end{equation}
\[
- \frac{1}{2}gp^{\mu}
( \{ ({\cal F}_{\mu \nu}^{T})_{L}, \frac{\partial f^{T(1)}}
{\partial p_{\nu}} \} -
\langle \{ ({\cal F}_{\mu \nu}^{T})_{L}, \frac{\partial f^{T(1)}}
{\partial p_{\nu}} \} \rangle) -
\frac{1}{2} g p^{\mu}
\{ ({\cal F}_{\mu \nu}^{T})_{NL} -
\langle ({\cal F}_{\mu \nu}^{T})_{NL} \rangle , \frac{\partial f^{R}}
{\partial p_{\nu}} \} ,
\]
\begin{equation}
p^{\mu} \partial_{\mu} f^{T(3)} = igp^{\mu}(
[{\cal A}_{\mu}^{T},f^{T(2)}] -
\langle [{\cal A}_{\mu}^{T},f^{T(2)}] \rangle) -
\frac{1}{2}gp^{\mu}
( \{ ({\cal F}_{\mu \nu}^{T})_{L}, \frac{\partial f^{T(2)}}
{\partial p_{\nu}} \}
\label{eq:a2}
\end{equation}
\[
- \langle \{ ({\cal F}_{\mu \nu}^{T})_{L}, \frac{\partial f^{T(2)}}
{\partial p_{\nu}} \} \rangle) -
\frac{1}{2}gp^{\mu}
( \{ ({\cal F}_{\mu \nu}^{T})_{NL}, \frac{\partial f^{T(1)}}
{\partial p_{\nu}} \} -
\langle \{ ({\cal F}_{\mu \nu}^{T})_{NL}, \frac{\partial f^{T(1)}}
{\partial p_{\nu}} \} \rangle ) \; etc..
\]

We rewrite the Yang-Mills equation (\ref{eq:w2}), connecting a gauge
field with a color current, in the following form
\begin{equation}
\partial_{\mu}(F^{T \mu \nu})_{L} - \xi^{-1} \partial^{\nu}
\partial^{\mu}A_{\mu}^{T} + j^{T(1) \nu} =
\label{eq:s2}
\end{equation}
\[
= -j_{NL}^{T \nu} +
ig \partial_{\mu}([A^{T \mu},A^{T \nu}] - \langle [A^{T \mu},
A^{T \nu}] \rangle )
\]
\[
+ ig([A^{T}_{\mu},(F^{T \mu \nu})_{L}] - \langle [A_{\mu}^{T},
(F^{T \mu \nu})_{L}] \rangle ) +
g^{2}([A_{\mu}^{T},[A^{T \mu},A^{T \nu}]] -
\langle [A_{\mu}^{T},[A^{T \mu},A^{T \nu}]] \rangle ) .
\]
Here, on the l.h.s. we collect all linear terms with
respect to $A_{\mu}^{T}$ and we denote: $j_{NL}^{T \nu} \equiv
j^{T(2) \nu} + j^{T(3) \nu} + \ldots.$
It will be shown that at leading order in $g$ only, the first two terms
in $j_{NL}^{T \nu}$ need to be kept.

It is not difficult to obtain the explicit form of the terms in the
color current expansion (\ref{eq:u2})
from the system of equations (\ref{eq:o2})-(\ref{eq:a2}) and the relation
(\ref{eq:i2}).
In the momentum representation and to the leading order in coupling constant
up to the third-order terms, we have [10]
\begin{equation}
j^{T(1) \mu}(k)= \Pi^{\mu \nu}(k)A^{T}_{\nu}(k),
\label{eq:d2}
\end{equation}
where
\begin{equation}
\Pi^{\mu \nu}(k)=g^{2}N_c \int \, d^{4}p \, \frac{p^{\mu}(p^{\nu}(k
\partial_{p})-(kp) \partial^{\nu}_{p})f^0}{pk + i p_{0} \epsilon}, \;
\epsilon \rightarrow + 0
\label{eq:f2}
\end{equation}
is the high temperature polarization tensor;
\begin{equation}
j^{T(2)a \mu}(k)= f^{abc} \int \, S^{(II) \mu \nu \lambda}_{k,k_{1},k_{2}}
(A^{b}_{\nu}(k_{1})A_{\lambda}^{c}(k_{2})-
\langle A^{b}_{\nu}(k_{1})A_{\lambda}^{c}(k_{2}) \rangle)
\delta (k - k_{1} - k_{2}) dk_{1}dk_{2},
\label{eq:g2}
\end{equation}
where
\begin{equation}
S_{k,k_{1},k_{2}}^{(II) \mu \nu \lambda}=
-ig^{3}N_c \int \, d^{4}p \, \frac{p^{\mu}p^{\nu}p^{\lambda}}
{pk + ip_{0} \epsilon} \, \frac {(k_{2} \partial_{p}f^0)}
{pk_{2} + ip_{0} \epsilon};
\label{eq:h2}
\end{equation}
\[
j^{T(3)a \mu}(k)= f^{abf}f^{fde}  \int \, \Sigma_{k,k_{1},k_{2},k_{3}}^{(II)
\mu \nu \lambda \sigma}(
A_{\nu}^{b}(k_{3})A_{\lambda}^{d}(k_{1})A_{\sigma}^{e}(k_{2})-
A_{\nu}^{b}(k_{3}) \langle A_{\lambda}^{d}(k_{1})
A_{\sigma}^{e}(k_{2}) \rangle
\]
\begin{equation}
- \langle A_{\nu}^{b}(k_{3})A_{\lambda}^{d}(k_{1})A_{\sigma}^{e}(k_{2}) \rangle )
\delta (k - k_{1} - k_{2} - k_{3}) \, dk_{1}dk_{2}dk_{3},
\label{eq:j2}
\end{equation}
and
\begin{equation}
\Sigma_{k,k_{1},k_{2},k_{3}}^{(II) \mu \nu \lambda \sigma} =
-g^{4}N_c \int \, d^{4}p \,
\frac{p^{\mu}p^{\nu}p^{\lambda}p^{\sigma}}{pk + ip_{0} \epsilon}
\, \frac{1}{p(k_{1} +k_{2}) + ip_{0} \epsilon}
\, \frac{(k_{2} \partial_p f^0)}{pk_{2} + ip_{0} \epsilon} .
\label{eq:k2}
\end{equation}
For simplicity, hereafter we drop the superscript $T$ of a gauge field.

Furthermore, we rewrite the equation (\ref{eq:s2}) in the momentum representation.
By inserting the linear part of the random current (\ref{eq:d2}) and
nonlinear corrections (\ref{eq:g2}) and (\ref{eq:j2}) into Eq. (\ref{eq:s2}),
one finds
\[
[k^{2}g^{\mu \nu} - (1+ \xi^{-1})k^{\mu}k^{\nu}- \Pi^{\mu \nu}(k)]
A^{b}_{\nu}(k)
\]
\begin{equation}
= f^{bcd} \int S^{\mu \nu \lambda}_{k,k_{1},k_{2}}
(A^{c}_{\nu}(k_{1})A_{\lambda}^{d}(k_{2})-
\langle A_{\nu}^{c}(k_{1})A_{\lambda}^{d}(k_{2}) \rangle)
\delta(k-k_{1}-k_{2}) dk_{1}dk_{2}
\label{eq:l2}
\end{equation}
\[
+ f^{bcf}f^{fde} \int \Sigma^{\mu \nu \lambda \sigma}
_{k,k_{1},k_{2},k_{3}}
(A^{c}_{\nu}(k_{3})A_{\lambda}^{d}(k_{1})A_{\sigma}^{e}(k_{2})-
A_{\nu}^{c}(k_{3}) \langle A_{\lambda}^{d}(k_{1})A_{\sigma}^{e}(k_{2}) \rangle -
\langle A_{\nu}^{c}(k_{3})A_{\lambda}^{d}(k_{1})A_{\sigma}^{e}(k_{2}) \rangle)
\]
\[
\times \delta(k-k_{1}-k_{2}-k_{3}) \, dk_{1}dk_{2}dk_{3}.
\]
Here,
$S^{\mu \nu \lambda}_{k,k_1,k_2} \equiv
S^{(I) \mu \nu \lambda}_{k,k_1,k_2} + S^{(II) \mu \nu \lambda}_{k,k_1,k_2},
\; \Sigma^{\mu \nu \lambda \sigma}_{k,k_1,k_2,k_3} \equiv
\Sigma^{(I) \mu \nu \lambda \sigma}_{k,k_1,k_2,k_3}
+ \Sigma^{(II) \mu \nu \lambda \sigma}_{k,k_1,k_2,k_3}$.
The functions $S^{(II)}$ and $\Sigma^{(II)}$ are defined by expressions
(\ref{eq:h2}) and (\ref{eq:k2}), respectively, and
\begin{equation}
S^{(I) \mu \nu \lambda}_{k,k_{1},k_{2}}=- ig(k^{\nu}
g^{\mu \lambda}+ k_{2}^{\nu}g^{\mu \lambda}- k_{2}^{\mu}g^{\nu \lambda}),
\; \Sigma_{k,k_{1},k_{2},k_{3}}^{(I) \mu \nu \lambda \sigma}=
g^{2}g^{\nu \lambda}g^{\mu \sigma}.
\label{eq:z2}
\end{equation}
These tensor structures are caused by self-action of a gauge field.
They are defined by nonlinear terms on the r.h.s. of equation (\ref{eq:s2}),
which are not associated with a color current in QGP. In section 5 equation
(\ref{eq:l2}) will be formally solved by iteration.

In [10] we introduce the correlation function of random oscillations
\begin{equation}
I_{\mu \nu}^{ab}(k^{\prime},k)= \langle A_{\mu}^{\dagger a}(k^{\prime})
A_{\nu}^{b}(k) \rangle.
\label{eq:x2}
\end{equation}
In order not to over burden equations by the symbol $" \ast"$ we use a dagger
$" \dagger"$ to denote complex conjugation.
In thermal equilibrium,
when the correlation function (\ref{eq:x2}) in the coordinate representation
depends only on the relative coordinates and time $ \triangle X=X^{\prime}
- X$, we have
\begin{equation}
I_{\mu \nu}^{ab}(k^{\prime},k)=I_{\mu \nu}^{ab}(k^{\prime})
\delta (k^{\prime}-k).
\label{eq:c2}
\end{equation}

Off-equilibrium perturbations which are slowly varying in space and time
lead to a delta function
broadering, and $I_{\mu \nu}^{ab}$ depends on both arguments $k$ and $k^{\prime}$.

Let us introduce $I_{\mu \nu}^{ab}(k^{\prime},k)=I_{\mu \nu}^{ab}
(k, \triangle k)$, $\triangle k = k^{\prime} - k$ with
$\mid\triangle k / k\mid \ll 1$
and insert the spectral intensity function in the Wigner form
\[
I_{\mu \nu}^{ab}(k,x)= \int I_{\mu \nu}^{ab}(k, \triangle k)
{\rm e}^{- i \triangle kx} d \triangle k,
\]
depending slowly on $x$.

Now we multiple equation (\ref{eq:l2}) by
$A^{\dagger a}_{\mu}(k^{\prime})$, subtract to it the complex-conjugated equation
(with the replacement $k \leftrightarrow k^{\prime} \;,
\; a \leftrightarrow b $) and average the equation using the formula
(\ref{eq:x2}). Furthermore we expand the polarization tensor into Hermitian and anti-Hermitian
parts
\[
\Pi^{\nu \sigma}(k)= \Pi^{H \nu \sigma}(k) + \Pi^{A \nu \sigma}(k) \;,
\; \Pi^{H \nu \sigma}(k)= \Pi^{\dagger H \sigma \nu}(k) \;, \;
\Pi^{A \nu \sigma}(k)= - \Pi^{\dagger A \sigma \nu}(k).
\]

The term with $ \Pi^{A}$ corresponds to linear Landau damping.
As was shown by Heinz and Siemens [33], that linear Landau damping for waves
in QGP is absent and hence, this term vanishes.
We expand the remaining terms on the l.h.s.
in powers of $ \triangle k$ and keep only linear ones.
This corresponds to ${\it a gradient \, expansion}$ procedure, usually used
in the derivation of kinetic equations.
Multiplying the difference equation by ${\rm e}^{- i \triangle kx}$
and integrating over $ \triangle k$ with regard to
\[
\int \triangle k_{\lambda} \, I_{\mu \nu}^{ab}(k, \triangle k)
\, {\rm e}^{- i \triangle kx} d \triangle k=
i \frac{\partial I_{\mu \nu}^{ab}(k,x)}{\partial x^{\lambda}},
\]
we finally obtain the equation, which is a starting point for our further research
\begin{equation}
\frac{\partial}{\partial k_{\lambda}}[k^{2}g^{\mu \nu}
-(1+ \xi^{-1})k^{\mu}k^{\nu} -
\Pi^{H \mu \nu}(k)] \frac{\partial I_{\mu \nu}^{ab}}{\partial x^{\lambda}} =
\label{eq:v2}
\end{equation}
\[
= - i \int \, dk^{\prime} \{f^{bcd} \,
S_{k,k_{1},k_{2}}^{\mu \nu \lambda} \langle
A_{\mu}^{\dagger a}(k^{\prime})A_{\nu}^{c}(k_{1})A_{\lambda}^{d}
(k_{2}) \rangle \delta (k-k_{1}-k_{2}) dk_1dk_2
\]
\[
+ f^{bcf}f^{fde} \Sigma_{k,k_{1},k_{2},k_{3}}^{\mu \nu \lambda
\sigma} (\langle A_{\mu}^{\dagger a}(k^{\prime})A_{\nu}^{c}(k_{3})
A_{\lambda}^{d}(k_{1})A_{\sigma}^{e}(k_{2}) \rangle -
\langle A_{\mu}^{\dagger a}(k^{\prime})A_{\nu}^{c}(k_{3}) \rangle
\langle A_{\lambda}^{d}(k_{1})A_{\sigma}^{e}(k_{2}) \rangle)
\]
\[
\times \delta (k-k_{1}-k_{2}-k_{3}) dk_{1}dk_{2}dk_{3} -
(a \leftrightarrow b, \, k \leftrightarrow k^{\prime}, \, compl. \,conj.) \}.
\]

\section{\bf Consistency with gauge symmetry}
\setcounter{equation}{0}

In this section we shall discuss the consistency of the approximation scheme
which we use with the requirement of the non-Abelian gauge symmetry.

The initial dynamical equation (\ref{eq:r2}) and the Yang-Mills equation (\ref{eq:w2})
(without the gauge-fixing condition) transform covariantly under local
transformations
\[
\bar{A}_{\mu} (X) = h (X) ( A_{\mu} (X) + \frac{i}{g} \partial_{\mu} )
h^{\dagger} (X) , \;
h (X) = \exp \, (i \theta^a (X) t^a)
\]
with the parameter $\theta^a (X)$. We also have transformation of gluon
distribution function [13]
\[
\bar{f}(p, X) = H (X) f(p, X) H^{\dagger} (X) ,
\]
where $H^{ab} (X) = {\rm Sp} [t^a h(X) t^b h^{\dagger} (X)]$.

As is known (see, e.g. [17]), after the splitting (\ref{eq:rr2}), (\ref{eq:rrr2})
the resulting equations left two symmetries: the ${\it background \; gauge \;
symmetry}$,
\begin{equation}
\bar{A}_{\mu}^{R} (X) = h (X) ( A_{\mu}^{R} (X) + \frac{i}{g} \partial_{\mu} )
h^{\dagger} (X) , \;
\bar{A}_{\mu}^{T} (X) = h (X) A_{\mu}^{T} (X) h^{\dagger} (X) ,
\label{eq:qw}
\end{equation}
and the ${\it fluctuation \; gauge \; symmetry}$,
\begin{equation}
\bar{A}_{\mu}^{R} (X) = 0 , \;
\bar{A}_{\mu}^{T} (X) = h (X) ( A_{\mu}^{R} (X) + A_{\mu}^{T} (X) +
\frac{i}{g} \partial_{\mu} )
h^{\dagger} (X) .
\label{eq:qe}
\end{equation}

The condition which we impose on a regular part of the gauge field $A_{\mu}^{R}$ in
the preceding section and the
requirement that the statistical average of the fluctuation vanishes
$\langle A_{\mu}^{T} \rangle = 0$, break down both types of symmetry (\ref{eq:qw})
and (\ref{eq:qe}). Thus in the case of a gauge transformation (\ref{eq:qw})
we obtain $\bar{A}_{\mu}^{R} \neq 0$, and in the case of (\ref{eq:qe})
we arrive at non-invariance of the constraint
$\langle A_{\mu}^{T} \rangle = 0$.
Moreover, the introduced correlation function (\ref{eq:x2}) also has an explicitly
gauge non-covariant character. This leads to the fact that
calculations in the previous section are gauge non-covariant, and therefore
the value of these manipulations is doubtful.

Nevertheless, there is a special case, when the preceding (and following)
conclusions are justified. This is a case of colorless fluctuation,
where $I^{ab}_{\mu \nu}(k,x)= \delta^{ab}I_{\mu \nu}(k,x)$.
We can obtain a gauge-invariant equation for $I_{\mu \nu}(k,x)$ only
in this restriction, in spite of the fact that the intermediate calculations
spoil non-Abelian gauge symmetry of the initial equations
(\ref{eq:w2})-(\ref{eq:r2}).

In principle, we shall be  able to maintain an explicit
background gauge symmetry (\ref{eq:qw}) at each step of our calculations,
as has been done, for example, by Blaizot and Iancu [18] for derivation of
the Boltzmann equation describing the relaxation of ultrasoft color excitations.
First of all we assume that
$A^{R}_{\mu} \neq 0$. Then as the gauge-fixing condition for the random field
$A^{T}_{\mu}$, we choose the background field gauge
\begin{equation}
{\cal D}^{R}_{\mu}(X) A^{T \, \mu}(X) =0, \;
{\cal D}^{R}_{\mu}(X) \equiv \partial_{\mu} - igA^{R}_{\mu}(X),
\label{eq:qr}
\end{equation}
which is manifestly covariant with respect to gauge transformations
of the background gauge field $A^{R}_{\mu}(X)$. Lastly we define
a gauge-covariant Wigner function as in $[13, \,18]$
\[
\acute{I}^{ab}_{\mu \nu}(k,x)=
\int \, \acute{I}^{ab}_{\mu \nu}(s,x) \, {\rm e}^{iks} {\rm d}s, \;
s \equiv X_1 - X_2, \, x \equiv \frac{1}{2}(X_1 + X_2),
\]
where
\[
\acute{I}^{ab}_{\mu \nu}(s,x) \equiv
U^{a a^{\prime}}(x,x + \frac{s}{2}) \,
I^{a^{\prime}b^{\prime}}(x + \frac{s}{2},x - \frac{s}{2}) \,
U^{b^{\prime}b}(x - \frac{s}{2},x),
\]
instead of the usual Wigner function $I_{\mu \nu}^{a b} (k, x)$,
whose `poor' transformation properties follow from the initial definition
$I^{ab}_{\mu \nu}(X_1,X_2) = \langle
A^{T a}_{\mu}(X_1)A^{T b}_{\nu}(X_2) \rangle$.
The function $U(x,y)$ is the non-Abelian parallel transporter
\[
U(x,y) = {\rm P}
\exp \Big\{ -ig \int_{\gamma} {\rm d}z^{\mu} A^{R}_{\mu}(z) \Big\}.
\]
The path $\gamma$ is the straight line joining $x$ and $y$.

The derivation of the kinetic equation for plasmons in this approach
becomes quite cumbersome and non-trivial. For example, on the l.h.s. of the
equations for a random part of the distribution
(\ref{eq:o2})-(\ref{eq:a2}), the covariant derivative ${\cal D}^{R}_{\mu}$
will be used instead of the ordinary one $\partial_{\mu}$.
Besides, we cannot assume that the regular part of the distribution function
is specified and equal to the Bose-Einstein distribution (\ref{eq:t2}). It is
necessary to also take
into account their change using the kinetic equation (\ref{eq:tt2}) with
the collision terms on the r.h.s. of (\ref{eq:tt2}), which describes the
backreaction of the background distributions from the soft fluctuations.
The correlators on the r.h.s. of equation (\ref{eq:tt2}) can be expressed in terms
of the function $\acute{I}_{\mu \nu}^{a b}$ and the distribution of hard transverse
gluons $f^{R} (p, X)$ only.

However, if we restrict our consideration to the study of colourless
excitations and replace
the distribution function of hard gluons by its equilibrium
value (\ref{eq:t2}), then this leads to an effective vanishing of terms
with mean field $A^{R}_{\mu}$. This follows, for example, from the analysis of
the derivation of the Boltzmann equation by Blaizot and Iancu [18].
Therefore, the simplest way to derive the kinetic equations for
soft colorless QGP excitations is to assume $A^{R}_{\mu} = 0$
and to use the prime gauge non-covariant correlator (\ref{eq:x2}).
In this case the background field gauge (\ref{eq:qr}) is reduced to
a covariant one.
The resulting Boltzmann equation for colorless plasmons will be gauge invariant if all contributions
to the probability of plasmon-plasmon scattering at the leading order in
$g$ are taken into account (see also discussion in conclusion).

\section{\bf The interacting fields as the functions
of free fields}
\setcounter{equation}{0}

Let us define approximate solution of equation (\ref{eq:l2}) accurate up to
third-order in the oscillations amplitude of the free field. For this purpose, it is
convenient to write this equation in a more compact form.

We introduce the following notation
\begin{equation}
J^{T(2)a \mu}(k) \equiv  f^{abc} \int \, S^{\mu \nu \lambda}_{k,k_{1},k_{2}}
(A^{b}_{\nu}(k_{1})A_{\lambda}^{c}(k_{2})-
\langle A^{b}_{\nu}(k_{1})A_{\lambda}^{c}(k_{2}) \rangle)
\delta (k - k_{1} - k_{2}) dk_{1}dk_{2},
\label{eq:q3}
\end{equation}
\[
J^{T(3)a \mu}(k) \equiv f^{abf}f^{fde}  \int \, \Sigma_{k,k_{1},k_{2},k_{3}}^{
\mu \nu \lambda \sigma}(
A_{\nu}^{b}(k_{3})A_{\lambda}^{d}(k_{1})A_{\sigma}^{e}(k_{2})-
A_{\nu}^{b}(k_{3}) \langle A_{\lambda}^{d}(k_{1})
A_{\sigma}^{e}(k_{2}) \rangle
\]
\begin{equation}
- \langle A_{\nu}^{b}(k_{3})A_{\lambda}^{d}(k_{1})A_{\sigma}^{e}(k_{2}) \rangle )
\delta (k - k_{1} - k_{2} - k_{3}) \, dk_{1}dk_{2}dk_{3}.
\label{eq:w3}
\end{equation}
The expressions (\ref{eq:q3}) and (\ref{eq:w3}) present the nonlinear color
currents, including the self-action effects of gauge fields, in contrast to
(\ref{eq:g2}) and (\ref{eq:j2}).

Using $\,^{\ast}{\cal D}_{\mu \nu}(k)$ we denote the medium modified (retarded)
gluon propagator, which in a covariant gauge has a form
\begin{equation}
\,^{\ast}{\cal D}_{\mu \nu}(k)= -
P_{\mu \nu}(k) \,^{\ast} \Delta^t(k) -
Q_{\mu \nu}(k) \,^{\ast} \Delta^l(k) +
\xi D_{\mu \nu}(k) \Delta^0(k),
\label{eq:e3}
\end{equation}
where $\,^{\ast} \Delta^{t,l}(k) = 1/(k^2 - \Pi^{t,l}(k)), \,
\Pi^t(k) = \frac{1}{2} \Pi^{\mu \nu}(k) P_{\mu \nu}(k), \,
\Pi^l(k) = \Pi^{\mu \nu}(k) Q_{\mu \nu}(k); \,
\Delta^0(k) = 1/k^2$.
The Lorentz matrices in (\ref{eq:e3}) are members of the basis
\[
P_{\mu \nu} (k) = g_{\mu \nu} -
D_{\mu \nu}(k) -
Q_{\mu \nu}(k), \,
Q_{\mu \nu}(k) =
\frac{\bar{u}_{\mu} (k) \bar{u}_{\nu} (k)}{\bar{u}^2(k)}, \,
C_{\mu \nu} (k) =
- \frac{(\bar{u}_{\mu} (k) k_{\nu} +
\bar{u}_{\nu} (k) k_{\mu})}{\sqrt{- 2 k^2 \bar{u}^2(k)}},
\]
\begin{equation}
D_{\mu \nu} = k_{\mu} k_{\nu}/k^2, \,
\bar{u}_{\mu} (k) = k^2 u_{\mu} - k_{\mu} (k u).
\label{eq:r3}
\end{equation}
Let us assume that we are in the rest frame of a heat bath, so that
$u_{\mu}=(1,0,0,0).$

Using the above introduced functions, the equation (\ref{eq:l2}) can be rewritten
in the form
\begin{equation}
\,^{\ast}{\cal D}^{-1 \, \mu \nu}(k) A^{a}_{\nu}(k) =
-J^{T(2)a \mu}(A,A) - J^{T(3)a \mu}(A,A,A).
\label{eq:t3}
\end{equation}
The nonlinear integral equation (\ref{eq:t3}) is solved by the approximation
scheme method - ${\it the weak \, free \, field \, expansion}$ (small perturbations).
Discarding the nonlinear terms in $A$ on the r.h.s. of equation (\ref{eq:t3}),
we obtain in the first approximation
\[
\,^{\ast}{\cal D}^{-1 \, \mu \nu}(k) A^{a}_{\nu}(k) = 0.
\]
The solution of this equation, which we denote by $A^{(0)a}_{\mu}(k)$
is the solution for free fields.

Further keeping the term, quadratic in the field on
the r.h.s. of Eq. (\ref{eq:t3}),
we derive the equation
\[
\,^{\ast}{\cal D}^{-1 \, \mu \nu}(k) A^{a}_{\nu}(k) =
-J^{T(2)a \mu}(A^{(0)},A^{(0)}),
\]
where on the r.h.s. we substitute free fields instead of interacting ones.
The general solution of the last equation can be given in the form
\[
A^{a}_{\mu}(k) = A^{(0)a}_{\mu}(k)
- \,^{\ast}{\cal D}_{\mu \nu}(k) J^{T(2)a \nu}(A^{(0)},A^{(0)}).
\]
This approximate solution was used in research of nonlinear plasmon damping
in QGP [10].

The following term in the expansion of interacting fields is defined from
equation
\[
\,^{\ast}{\cal D}^{-1 \, \mu \nu}(k) A^{a}_{\nu}(k) =
-J^{T(2)a \mu}(- \,^{\ast}{\cal D} J^{T(2)}(A^{(0)},A^{(0)}),A^{(0)})-
J^{T(2)a \mu}(A^{(0)},- \,^{\ast}{\cal D} J^{T(2)}(A^{(0)},A^{(0)}))
\]
\begin{equation}
- J^{T(3)a \mu}(A^{(0)},A^{(0)},A^{(0)}).
\label{eq:y3}
\end{equation}
Using the explicit expressions for the currents (\ref{eq:q3}) and (\ref{eq:w3}),
after cumbersome algebraic transformations, we obtain the form of the interacting
field from the equation (\ref{eq:y3}) with the accuracy required for our further calculations
\begin{equation}
A^{a}_{\mu}(k) = A^{(0)a}_{\mu}(k)
- \,^{\ast}{\cal D}_{\mu \nu}(k) J^{T(2)a \nu}(A^{(0)},A^{(0)})-
\,^{\ast}{\cal D}_{\mu \nu}(k) \tilde{J}^{T(3)a \nu}(A^{(0)},A^{(0)},A^{(0)}).
\label{eq:u3}
\end{equation}
Here, the third-order color current on the r.h.s. is defined by the expression
\[
\tilde{J}^{T(3)a \nu}(A^{(0)},A^{(0)},A^{(0)}) \equiv
f^{abf}f^{fde}  \int \, \tilde{\Sigma}_{k,k_{1},k_{2},k_{3}}^{
\nu \lambda \sigma \rho}(
A_{\lambda}^{(0)b}(k_{3})A_{\sigma}^{(0)d}(k_{1})A_{\rho}^{(0)e}(k_{2})-
\]
\begin{equation}
- A_{\lambda}^{(0)b}(k_{3}) \langle A_{\sigma}^{(0)d}(k_{1})
A_{\rho}^{(0)e}(k_{2}) \rangle)
\delta (k - k_{1} - k_{2} - k_{3}) \, dk_{1}dk_{2}dk_{3},
\label{eq:i3}
\end{equation}
where
\begin{equation}
\tilde{\Sigma}_{k,k_{1},k_{2},k_{3}}^{\nu \lambda \sigma \rho}
\equiv \Sigma_{k,k_{1},k_{2},k_{3}}^{\nu \lambda \sigma \rho} -
\frac{1}{2} \,^{\ast}{\cal D}_{\delta \gamma}(k_1 + k_2)
(S_{k,k_{3},k_{1} + k_{2}}^{\nu \lambda \delta} -
S_{k,k_{1} + k_{2},k_{3}}^{\nu \delta \lambda})
(S_{k_{1} + k_{2},k_{1},k_{2}}^{\gamma \sigma \rho} -
S_{k_{1} + k_{2},k_{2},k_{1}}^{\gamma \rho \sigma}) ,
\label{eq:o3}
\end{equation}
and we take into account, that the third-order correlation function
$\langle A^{(0)} A^{(0)} A^{(0)} \rangle$ vanishes by virtue of the fact
that $A^{(0)}$ represents the amplitude fully non-correlative gauge fields.
The factor of $\frac{1}{2}$, in front of the second term on the r.h.s. of (\ref{eq:o3})
arises from symmetrization with respect to permutation of the potentials
$A^{(0)d}_{\sigma}(k_1)$ and $A^{(0)e}_{\rho}(k_2)$ in the expression
(\ref{eq:i3}). The current (\ref{eq:i3}) may be interpreted as a certain
third-order effective color current, in contrast to the initial `bare'
expression (\ref{eq:w3}).

\section{\bf The correspondence principle}
\setcounter{equation}{0}

For the determination of the probability of plasmon-plasmon scattering in a gluon plasma
the method developed in the theory of the nonlinear
processes in electron-ion plasma and known as {\it the correspondence principle}
[24, 30], is usable. For the non-Abelian plasma this approach is especially
effective in the temporal gauge, when we have closer correspondence
with the electrodynamics of an ordinary plasma. The gist of this method is as follows.

The change in the plasmon numbers, caused by spontaneous processes of
four-plasmon decays only, is
\begin{equation}
\left( \frac{\partial N_{\bf k}^{l}}{\partial t} +
{\bf V}_{\bf k}^{l} \, \frac{\partial N_{\bf k}^{l}}
{\partial {\bf x}} \right)^{sp}_4 =
\int \frac{{\rm d} {\bf k}_1}{(2 \pi)^3} \frac{{\rm d} {\bf k}_2}{(2 \pi)^3}
\frac{{\rm d} {\bf k}_3}{(2 \pi)^3}
\label{eq:q4}
\end{equation}
\[
\times (2 \pi)^4 \delta (\omega_{\bf k}^l + \omega_{{\bf k}_1}^l
- \omega_{{\bf k}_2}^l - \omega_{{\bf k}_3}^l)
\delta ({\bf k} + {\bf k}_1 - {\bf k}_2 - {\bf k}_3)
\, w ( {\bf k}, {\bf k}_1; {\bf k}_2, {\bf k}_3 )
N_{{\bf k}_1}^l N_{{\bf k}_2}^l N_{{\bf k}_3}^l.
\]
This equation follows from equation (\ref{eq:n1}) in the limit of small
intensity $N_{\bf k}^l \rightarrow 0$. In this case the change of energy of
the longitudinal excitations is
\begin{equation}
\left( \frac{{\rm d}{\cal E}}{{\rm d} t} \right)^{sp}_4 =
\int \frac{{\rm d}{\bf k}}{(2 \pi)^3} \frac{{\rm d} {\bf k}_1}{(2 \pi)^3}
\frac{{\rm d} {\bf k}_2}{(2 \pi)^3}   \frac{{\rm d} {\bf k}_3}{(2 \pi)^3}
\label{eq:qw4}
\end{equation}
\[
\times (2 \pi)^4 \delta (\omega_{\bf k}^l + \omega_{{\bf k}_1}^l
- \omega_{{\bf k}_2}^l - \omega_{{\bf k}_3}^l)
\delta ({\bf k} + {\bf k}_1 - {\bf k}_2 - {\bf k}_3) \, \omega_{\bf k}^l
w ( {\bf k}, {\bf k}_1; {\bf k}_2, {\bf k}_3 )
N_{{\bf k}_1}^l N_{{\bf k}_2}^l N_{{\bf k}_3}^l.
\]
On the other hand the value $({\rm d}{\cal E}/{\rm d}t)_4^{sp}$ represents the
emitted radiant power of the longitudinal waves ${\cal I}^l$, which in turn is
equal to the work done by the radiation field with the color current, creating it, in unit time
\begin{equation}
{\cal I}^l = \int {\rm d}{\bf x} \, \langle {\bf E}^a ({\bf x},t) \,
{\bf J}^a ({\bf x},t) \rangle =
\int \, \frac{{\rm d} \omega}{2 \pi} \frac{{\rm d} \omega^{\prime}}{2 \pi}
\frac{{\rm d}{\bf k}}{(2 \pi)^3} \,
\langle {\bf E}^a_{{\bf k}, \omega^{\prime}} \,
{\bf J}^a_{-{\bf k},- \omega} \rangle {\rm e}^{-i( \omega^{\prime} -
\omega)t}=
\label{eq:qq4}
\end{equation}
\[
=  \frac{1}{2} \int \, \frac{{\rm d} \omega}{2 \pi} \frac{{\rm d}
\omega^{\prime}}{2 \pi} \frac{{\rm d}{\bf k}}{(2 \pi)^3}
\, i \left( \frac{1}{\omega^{\prime} \varepsilon^l (\omega^{\prime}, {\bf k})}
- \frac{1}{\omega \varepsilon^{\dagger l} (\omega, {\bf k})} \right)
\, \frac{k^i k^j}{{\bf k}^2}
\langle J_{{\bf k}, \omega}^{\dagger a i} \,
 J_{{\bf k}, \omega^{\prime}}^{a j} \rangle
{\rm e}^{- i (\omega^{\prime} - \omega) t}.
\]
Here, $E^{ai}({\bf x},t) = - \partial A^{ai}({\bf x},t)/ \partial t$
is chromoelectric field in the temporal gauge.
The sign on the r.h.s. of (\ref{eq:qq4}) corresponds to the choice of sign in
front of the current in the Yang-Mills equation (\ref{eq:w2}). In conclusion
of the last line (\ref{eq:qq4}) we take into account that the Fourier-component of a field
${\bf E}_{{\bf k}, \omega}^a = {\bf k} E_{{\bf k}, \omega}^a / \vert {\bf k}
\vert$ is associated with ${\bf J}_{{\bf k}, \omega}^a$ by the Yang-Mills equation
\[
E_{{\bf k}, \omega}^a = \frac{i}
{\omega \varepsilon^l ( \omega, {\bf k} )}
\; \frac{({\bf k} \cdot {\bf J}_{{\bf k}, \omega}^a)}
{\vert {\bf k} \vert} .
\]
In order to define the probability of the four-plasmon decays,
the correlation function on the r.h.s. of equation (\ref{eq:qq4}) has to contain
terms of six-order in the free field $A^{(0)}$.
The required sixth-order correlator yields the color current
$J^{T (3) ai}$ (\ref{eq:w3}) (more precisely, its expression in
the temporal gauge). However here, it is also necessary to take into account
the effects, that arise from iteration of the current $J^{T(2) ai}$
(\ref{eq:q3}). Defining in this way all necessary contributions,
making the correlation decoupling of the sixth-order correlators
in terms of pairs and next expressing next
$\langle A^{(0)} A^{(0)} \rangle$ in terms of $N^l$,
we obtain an emitted radiant power ${\cal I}^l$. Comparing
${\cal I}^l$ with (\ref{eq:qw4}),
one identifies the required probability $w({\bf k}, {\bf k}_1; {\bf k}_2, {\bf k}_3)$.

However this method encountered certain difficulties in deciding on the other
gauges, e.g. the covariant gauge. Here, it is convenient for the definition of
the probability of plasmon-plasmon scattering to start immediately from
equation (\ref{eq:v2}). Using the above-obtained expression (\ref{eq:u3}) for
the potentials of interacting fields, by a simple search we extract all
the sixth-order correlators responsible for four-plasmon
decays of the type (\ref{eq:c1}). The rule used to choose the relevant terms is defined
by the simple fact that when we make the correlation decoupling of the sixth-order correlators
in the terms of the pairs (in order to define the product
$N_{{\bf k}_1}^l N_{{\bf k}_2}^l N_{{\bf k}_3}^l$) as a factor, the
$\delta$-functions arise in the form
\begin{equation}
\delta (\omega_{\bf k}^l + \omega_{{\bf k}_1}^l
- \omega_{{\bf k}_2}^l - \omega_{{\bf k}_3}^l)
\delta ({\bf k} + {\bf k}_1 - {\bf k}_2 - {\bf k}_3).
\label{eq:qqq4}
\end{equation}
Also the coefficient function preceding
$N_{{\bf k}_1}^l N_{{\bf k}_2}^l N_{{\bf k}_3}^l$ must satisfy
properties (\ref{eq:m1}). As will be shown below, these conditions are
sufficient to calculate the plasmon-plasmon scattering probability in the
covariant gauge (this rather cumbersome and physically not quite
transparent approach is suited for other gauges).

At first, we consider the contribution to the r.h.s. of the initial
equation (\ref{eq:v2}), associated with $\Sigma$-functions. Here, for convenience of
further reference we write it separately
\[
- i \int \, dk^{\prime} \{f^{bcf} f^{fde} \,
\Sigma_{k,k_{1},k_{2}, k_{3}}^{\mu \nu \lambda \sigma} ( \langle
A_{\mu}^{\dagger a}(k^{\prime})A_{\nu}^{c}(k_{3})A_{\lambda}^{d}
(k_{1}) A_{\sigma}^e (k_{2}) \rangle -
\langle A_{\mu}^{\dagger a}(k^{\prime})A_{\nu}^{c}(k_{3}) \rangle
\langle A_{\lambda}^{d}(k_{1})A_{\sigma}^{e}(k_{2}) \rangle )
\]
\begin{equation}
\times \delta (k-k_{1}-k_{2}-k_{3}) dk_{1}dk_{2}dk_{3} -
(a \leftrightarrow b, \, k \leftrightarrow k^{\prime}, \, compl. \, conj.) \}.
\label{eq:w4}
\end{equation}
One can obtain the sixth-order terms in the free field by two waves.
The first one is as follows. We substitute in turn the expression
from the r.h.s. of (\ref{eq:u3}), which contains only cubic terms
in the potentials of free fields, instead of each potential
of the interacting ones, i.e.
\[
A_{\mu}^a(k)  \rightarrow - \,^{\ast}{\cal D}_{\mu \mu^{\prime}} (k)
\tilde{J}^{T (3) a \mu^{\prime}} (A^{(0)}, A^{(0)}, A^{(0)}).
\]
We replace the remaining potentials by the rule $A_{\nu}^c(k_3) \rightarrow
A_{\nu}^{(0) c} (k_3)$, etc.
The second way is to substitute the quadratic term in $A^{(0)}$ from the
r.h.s. of (\ref{eq:u3}) instead of any two potentials of the interacting
fields, i.e.
\[
A_{\mu}^a(k)  \rightarrow - \,^{\ast}{\cal D}_{\mu \mu^{\prime}} (k)
{J}^{T (2) a \mu^{\prime}} (A^{(0)}, A^{(0)}).
\]
We replace the remaining potentials by free ones.
It is necessary to look at all possible substitutions in both first and
second ways.

The number of terms appearing can be cut if we note, that
it is need to keep only such terms in intermediate expressions which contain
the propagators $\,^{\ast}{\cal D}_{\mu \mu^{\prime}} (k)$ and
$\,^{\ast}{\cal D}_{\mu \mu^{\prime}} (k^{\prime})$.
These propagators give the terms, proportional to
\begin{equation}
\delta ( {\rm Re} \, \varepsilon^l(k)) =
\bigg( \frac{\partial {\rm Re} \, \varepsilon^l (k)}
{\partial \omega} \bigg)^{- 1}_{\omega = \omega_{\bf k}^l}
[ \delta ( \omega - \omega_{\bf k}^l ) +
\delta ( \omega + \omega_{\bf k}^l ) ] ,
\label{eq:e4}
\end{equation}
i.e. the factor taken into account the existence of plasmons with
wavevector ${\bf k}$ and energy $\omega_{\bf k}^l$, in spite of the fact
that the number density of the plasmon $N_{\bf k}^l$ is explicitly
absent.

Hence it follows that for the first way only the
replacement in (\ref{eq:w4})
\[
A_{\mu}^{\dagger a} (k^{\prime})  \rightarrow -
\,^{\ast}{\cal D}^{\dagger}_{\mu \mu^{\prime}} (k^{\prime})
\tilde{J}^{\dagger T (3) a \mu^{\prime}} (A^{(0)}, A^{(0)}, A^{(0)}) ,
\; A_{\nu}^c (k_3) \rightarrow A_{\nu}^{(0) c} (k_{3}), \ldots
\]
gives desired contribution (similar for the conjugate term). This leads (\ref{eq:w4})
to the expression
\[
i \int \, dk^{\prime} \{ f^{bcf} f^{fde} f^{ac^{\prime}g}
f^{g d^{\prime}e^{\prime}}
\, \,^{\ast}{\cal D}_{\mu \mu^{\prime}}^{\dagger} (k^{\prime})
\, \Sigma_{k,k_{1},k_{2},k_3}^{\mu \nu \lambda \sigma}
\tilde{\Sigma}_{k^{\prime},k_{1}^{\prime},
k_{2}^{\prime},k_3^{\prime}}^{\dagger \mu^{\prime} \nu^{\prime}
\lambda^{\prime} \sigma^{\prime}}
( \langle A_{\nu^{\prime}}^{(0) \dagger c^{\prime}}(k_3^{\prime})
A_{\lambda^{\prime}}^{(0) \dagger d^{\prime}}(k_1^{\prime})
A_{\sigma^{\prime}}^{(0) \dagger e^{\prime}}(k_2^{\prime})
\]
\[
\times A_{\nu}^{(0) c}(k_3) A_{\lambda}^{(0) d}(k_1)
A_{\sigma}^{(0) e}(k_2) \rangle
- \langle A_{\nu^{\prime}}^{(0) \dagger c^{\prime}}(k_3^{\prime})
A_{\lambda^{\prime}}^{(0) \dagger d^{\prime}}(k_1^{\prime})
A_{\sigma^{\prime}}^{(0) e^{\prime}}(k_2^{\prime})
A_{\nu}^{(0) c}(k_3) \rangle
\langle A_{\lambda}^{(0) d}(k_1) A_{\sigma}^{(0) e}(k_2) \rangle
\]
\begin{equation}
- \langle A_{\nu^{\prime}}^{(0) \dagger c^{\prime}}(k_3^{\prime})
A_{\nu}^{(0) c}(k_3) A_{\lambda}^{(0) d}(k_1)
A_{\sigma}^{(0) e}(k_2) \rangle
\langle A_{\lambda^{\prime}}^{(0) \dagger d^{\prime}}(k_1^{\prime})
A_{\sigma^{\prime}}^{(0) \dagger e^{\prime}}(k_2^{\prime}) \rangle
\label{eq:r4}
\end{equation}
\[
+ \langle A_{\nu^{\prime}}^{(0) \dagger c^{\prime}}(k_3^{\prime})
A_{\nu}^{(0) c}(k_3) \rangle
\langle A_{\lambda^{\prime}}^{(0) \dagger d^{\prime}}(k_1^{\prime})
A_{\sigma^{\prime}}^{(0) \dagger e^{\prime}}(k_2^{\prime}) \rangle
\langle A_{\lambda}^{(0) d}(k_1)
A_{\sigma}^{(0) e}(k_2) \rangle )
\]
\[
\times \delta (k-k_{1}-k_{2}-k_{3})
\delta (k^{\prime}-k_{1}^{\prime}-k_{2}^{\prime}-k_{3}^{\prime})
\prod_{i=1}^{3}dk_{i}dk_{i}^{\prime} -
(a \leftrightarrow b, \, k \leftrightarrow k^{\prime}, \, compl. \,conj.) \}.
\]

In the second case, at first step it should be replaced by
\[
A_{\mu}^{\dagger a} (k^{\prime})  \rightarrow -
\,^{\ast}{\cal D}^{\dagger}_{\mu \mu^{\prime}} (k^{\prime})
J^{\dagger T (2) a \mu^{\prime}} (A^{(0)}, A^{(0)}).
\]
This gives
\[
i \int \, dk^{\prime} \{ f^{bcf} f^{fde} f^{ab^{\prime}c^{\prime}}
\, \,^{\ast}{\cal D}_{\mu \mu^{\prime}}^{\dagger} (k^{\prime}) \,
\Sigma_{k,k_{1},k_{2},k_3}^{\mu \nu \lambda \sigma}
S_{{k}^{\prime},k_{1}^{\prime},
k_{2}^{\prime}}^{\dagger \mu^{\prime} \nu^{\prime}
\lambda^{\prime}}
( \langle A_{\nu^{\prime}}^{(0) \dagger b^{\prime}}(k_1^{\prime})
A_{\lambda^{\prime}}^{(0) \dagger c^{\prime}}(k_2^{\prime})
A_{\nu}^{c}(k_3) A_{\lambda}^d (k_1)
A_{\sigma}^e (k_2) \rangle
\]
\[
- \langle A_{\nu^{\prime}}^{(0) \dagger b^{\prime}}(k_1^{\prime})
A_{\lambda^{\prime}}^{(0) \dagger c^{\prime}}(k_2^{\prime}) \rangle
\langle A_{\nu}^{c}(k_3) A_{\lambda}^d (k_1) A_{\sigma}^e (k_2) \rangle -
\langle A_{\nu^{\prime}}^{(0) \dagger b^{\prime}}(k_1^{\prime})
A_{\lambda^{\prime}}^{(0) \dagger c^{\prime}}(k_2^{\prime})
A_{\nu}^{c}(k_3) \rangle \langle A_{\lambda}^d (k_1) A_{\sigma}^e (k_2) \rangle
\]
\begin{equation}
+ \langle A_{\nu^{\prime}}^{(0) \dagger b^{\prime}}(k_1^{\prime})
A_{\lambda^{\prime}}^{(0) \dagger c^{\prime}}(k_2^{\prime}) \rangle
\langle A_{\nu}^{c}(k_3) \rangle
\langle A_{\lambda}^d (k_1) A_{\sigma}^e (k_2) \rangle )
\label{eq:t4}
\end{equation}
\[
\times \delta(k^{\prime} - k_{1}^{\prime} - k_{2}^{\prime})
\delta (k-k_{1}-k_{2}-k_{3}) dk_{1}dk_{2}dk_{3}dk_{1}^{\prime}dk_2^{\prime} +
(k \leftrightarrow k^{\prime}, \, a \leftrightarrow b, \, compl.\, conj.) \}.
\]
By virtue of the stochasticity of gauge fields, the last term inside the parentheses
vanishes. We replace the remaining potentials of the interacting fields by the
rules
\[
A_{\nu}^c (k_3) \rightarrow
- \,^{\ast}{\cal D}_{\nu \nu^{\prime}} (k_3) J^{T (2) c \nu^{\prime}}
(A^{(0)}, A^{(0)}), \; A_{\lambda}^d (k_1) \rightarrow A_{\lambda}^{(0) d}
(k_1), \; A_{\sigma}^e (k_2) \rightarrow A_{\sigma}^{(0) e} (k_2), \; etc. .
\]

By inspecting the expression (\ref{eq:t4}), one sees that, complex-conjugate
and non-conjugate amplitudes of free fields enter by a non-symmetric
fashion. As will be shown below, in this case $\delta$-functions
(\ref{eq:qqq4}) type expressing the energy and
momentum conservation laws in the plasmon-plasmon scattering have not arisen.

By virtue of the fact that we restrict our consideration to the derivation
of the kinetic equation for colorless excitations, i.e.
\begin{equation}
\langle A^{(0) d} A^{(0) e} \rangle \sim \delta^{d e} , \;
\langle A^{(0) \dagger d^{\prime}} A^{(0) \dagger e^{\prime}} \rangle
\sim \delta^{d^{\prime} e^{\prime}} ,
\label{eq:y4}
\end{equation}
all terms inside the parentheses of the expression (\ref{eq:r4}), excepting the first term
(the sixth-order correlator) vanish because $\delta^{dc}, \, \delta^{d^{\prime}c^{\prime}},
\ldots,$ are contracted with antisymmetric structure constants.
Further, it is necessary to decouple the averaging of six potentials into pairs.
We define below, what the correlations decoupling is responsible for in the
four-plasmon decay processes.
By using the definition of the correlation function (\ref{eq:x2}),
(\ref{eq:c2}), we have
\begin{equation}
\langle A_{\nu^{\prime}}^{(0) \dagger c^{\prime}}(k_3^{\prime})
A_{\lambda^{\prime}}^{(0) \dagger d^{\prime}}(k_1^{\prime})
A_{\sigma^{\prime}}^{(0) \dagger e^{\prime}}(k_2^{\prime})
A_{\nu}^{(0)c}(k_3) A_{\lambda}^{(0) d}(k_1)
A_{\sigma}^{(0) e} (k_2) \rangle
\label{eq:u4}
\end{equation}
\[
= \langle A_{\nu^{\prime}}^{(0) \dagger c^{\prime}} (k_3^{\prime})
A_{\lambda^{\prime}}^{(0) \dagger d^{\prime}}(k_1^{\prime}) \rangle
\langle A_{\sigma^{\prime}}^{(0) \dagger e^{\prime}}(k_2^{\prime})
 A_{\lambda}^{(0) d} (k_1) \rangle
\langle  A_{\nu}^{(0) c} (k_3) A_{\sigma}^{(0) e} (k_2) \rangle + \ldots
\]
\[
= I_{\nu^{\prime} \lambda^{\prime}}(k_3^{\prime}) \delta^{c^{\prime}d^{\prime}}
\delta(k_3^{\prime} + k_1^{\prime})
I_{\sigma^{\prime} \lambda}(k_2^{\prime}) \delta^{e^{\prime}d}
\delta(k_2^{\prime} - k_1) I_{\nu \sigma}(k_3) \delta^{ce}
\delta(k_3 + k_2) + \ldots \,.
\]

After substitution of the first term on the r.h.s. (\ref{eq:u4}) into
(\ref{eq:r4}) and performing integration over
$dk_1 dk_3 dk_2^{\prime} dk_3^{\prime}$ and elementary color algebra, we obtain
\[
\delta^{ab} \, N_c^2 \, (-i)
\!\!\int\!\{ \,^{\ast}{\cal D}_{\mu \mu^{\prime}}^{\dagger} (k)
\Sigma_{k,k,k_{2},-k_1}^{\mu \nu \lambda \sigma}
\tilde{\Sigma}_{k,k_1^{\prime},k,-k_{1}^{\prime}}^
{\dagger \mu^{\prime} \nu^{\prime}
\lambda^{\prime} \sigma^{\prime}}
\, I_{\nu^{\prime} \lambda^{\prime}}(-k_1^{\prime})
I_{\sigma^{\prime} \lambda}(k) I_{\nu \sigma}(-k_2)
dk_1^{\prime} dk_2 -
(compl. \, conj.) \}.
\]
As can be seen from the last expression, the required $\delta$-functions (\ref{eq:qqq4}) are
not appeared and therefore this expression is not associated with the
plasmon-plasmon scattering and it should be dropped. This is a general rule.
The decomposition of averaging of free field amplitudes into the correlators
containing the pair of complex-conjugate potentials or one of the
nonconjugate potentials between the inside of the angular brackets (statistical
averaging), does not give a contribution to the process of
interest to us. For this reason, it is necessary to fully drop all contribution
in the process of the plasmon-plasmon scattering defined by the expression (\ref{eq:t4}),
since making the correlation decoupling, the pair with complex-conjugate
amplitudes or without conjugate necessarily arises.
We write out decoupling of the sixth-order correlator, which gives a
contribution to equation (\ref{eq:q4}). Suppressing color and Lorentz indices and
employing a condensed notion, $A_1 \equiv A_{\mu}^a (k_1)$, we have
\[
\langle A_{3^{\prime}}^{\dagger}
A_{2^{\prime}}^{\dagger}
A_{1^{\prime}}^{\dagger}
A_{3} A_{2} A_{1} \rangle =
3 \{ \langle A_{3^{\prime}}^{\dagger} A_{3} \rangle
\langle A_{1^{\prime}}^{\dagger}  A_{1} \rangle
\langle A_{2^{\prime}}^{\dagger}  A_{2} \rangle +
\langle A_{3^{\prime}}^{\dagger} A_{3} \rangle
\langle A_{1^{\prime}}^{\dagger}  A_{2} \rangle
\langle A_{2^{\prime}}^{\dagger}  A_{1} \rangle
\]
\[
+ \langle A_{3^{\prime}}^{\dagger} A_{1} \rangle
\langle A_{1^{\prime}}^{\dagger}  A_{3} \rangle
\langle A_{3^{\prime}}^{\dagger}  A_{2} \rangle +
\langle A_{3^{\prime}}^{\dagger} A_{1} \rangle
\langle A_{1^{\prime}}^{\dagger}  A_{2} \rangle
\langle A_{2^{\prime}}^{\dagger}  A_{3} \rangle +
\langle A_{3^{\prime}}^{\dagger} A_{2} \rangle
\langle A_{1^{\prime}}^{\dagger}  A_{3} \rangle
\langle A_{2^{\prime}}^{\dagger}  A_{1} \rangle
\]
\begin{equation}
+ \langle A_{3^{\prime}}^{\dagger} A_{2} \rangle
\langle A_{1^{\prime}}^{\dagger}  A_{1} \rangle
\langle A_{2^{\prime}}^{\dagger}  A_{3} \rangle \} .
\label{eq:i4}
\end{equation}

Now we consider the terms with $S$-functions on the r.h.s. of equation (\ref{eq:v2})
and here, we also write them separately
\[
- i \int \, dk^{\prime} \{ f^{bcd}
S_{k,k_{1},k_{2}}^{\mu \nu \lambda} \langle
A_{\mu}^{\dagger a} (k^{\prime}) A_{\nu}^c (k_1)
A_{\lambda}^{d} (k_2) \rangle
\delta(k - k_1 - k_2) dk_1 dk_2
\]
\begin{equation}
- f^{acd}
S_{k^{\prime},k_{1},k_{2}}^{\dagger \mu \nu \lambda} \langle
A_{\mu}^{b} (k) A_{\nu}^{\dagger c} (k_1)
A_{\lambda}^{\dagger d} (k_2) \rangle
\delta (k^{\prime} - k_1 - k_2) dk_1 dk_2 \}.
\label{eq:o4}
\end{equation}

According to the previous discussion, at a first step it is necessary to perform
the replacement
\[
A_{\mu}^{\dagger a} (k^{\prime})  \rightarrow -
\,^{\ast}{\cal D}^{\dagger}_{\mu \mu^{\prime}}
(k^{\prime})
\tilde{J}^{\dagger T (3) a \mu^{\prime}} (A^{(0)}, A^{(0)}, A^{(0)}),
\]
\[
A_{\mu}^{b} (k)  \rightarrow - \,^{\ast}{\cal D}_{\mu \mu^{\prime}}
(k)
\tilde{J}^{T (3) b \mu^{\prime}} (A^{(0)}, A^{(0)}, A^{(0)}).
\]
Furthermore we consequently replace the remaining two potentials of the interacting fields in the
correlators (\ref{eq:o4}) by
$A_{\nu}^{c} (k_1)  \rightarrow - \,^{\ast}{\cal D}_{\nu \nu^{\prime}}
(k_1)
J^{T (2) c \nu^{\prime}} (A^{(0)}, A^{(0)}), \; A_{\lambda}^d (k_2) \rightarrow
A_{\lambda}^{(0) d} (k_2)$, etc.
This automatically leads to symmetry of contribution in
$A^{(0)}$ and $A^{\dagger (0)}$. As result we obtain
\[
\frac{i}{2} \int \, \{ \,^{\ast}{\cal D}_{\mu \mu^{\prime}}^{\dagger} (k^{\prime})
f^{bcf} f^{fde} f^{ac^{\prime}g}
f^{g d^{\prime}e^{\prime}}
(S_{k,k_{3},k_{1}+k_2}^{\mu \nu \rho} -
S_{k,k_1+k_2,k_3}^{\mu \rho \nu})
\tilde{\Sigma}_{k^{\prime},k_{1}^{\prime},
k_{2}^{\prime}, k_{3}^{\prime}}^{\dagger \mu^{\prime} \nu^{\prime}
\lambda^{\prime} \sigma^{\prime}} \,^{\ast}{\cal D}_{\rho \rho^{\prime}}
(k_1 + k_2)
\]
\[
\times (S_{k_{1}+k_2,k_1, k_2}^{\rho^{\prime} \lambda \sigma} -
S_{k_1+k_2,k_2,k_1}^{\rho^{\prime} \sigma \lambda})
( \langle A_{\nu^{\prime}}^{\dagger (0)c^{\prime}}(k_3^{\prime})
A_{\lambda^{\prime}}^{\dagger (0) d^{\prime}}(k_1^{\prime})
A_{\sigma^{\prime}}^{\dagger (0) e^{\prime}}(k_2^{\prime})
A_{\lambda}^{(0)d} (k_1) A_{\sigma}^{(0) e} (k_2)
A_{\nu}^{(0) c} (k_3) \rangle
\]
\begin{equation}
- \langle A_{\nu^{\prime}}^{\dagger (0)c^{\prime}}(k_3^{\prime})
A_{\lambda^{\prime}}^{\dagger (0) d^{\prime}}(k_1^{\prime})
A_{\sigma^{\prime}}^{\dagger (0) e^{\prime}}(k_2^{\prime})
A_{\nu}^{(0) c} (k_3) \rangle \langle A_{\lambda}^{(0) d} (k_1)
A_{\sigma}^{(0) e} (k_2) \rangle
\label{eq:p4}
\end{equation}
\[
- \langle A_{\nu^{\prime}}^{\dagger (0)c^{\prime}}(k_3^{\prime})
A_{\lambda}^{(0) d}(k_1)
A_{\sigma}^{(0) e}(k_2)
A_{\nu}^{(0)c} (k_3) \rangle
\langle A_{\lambda^{\prime}}^{(0) \dagger d^{\prime}} (k_1^{\prime})
A_{\sigma^{\prime}}^{(0) \dagger e^{\prime}} (k_2^{\prime}) \rangle )
\]
\[
\times \delta (k - k_1 - k_2 - k_3)
\delta (k^{\prime} - k_1^{\prime} - k_2^{\prime} - k_3^{\prime})
\prod_{i=1}^{3}
dk_i dk_i^{\prime} - (a \leftrightarrow b, \,
k \leftrightarrow k^{\prime}, \, compl. conj.) \} .
\]
Because of (\ref{eq:y4}), in the last expression it is necessary
to retain only the sixth-order correlator.
Finally adding (\ref{eq:r4}) and (\ref{eq:p4}), we obtain the following equation,
instead of (\ref{eq:v2})
\[
\frac{\partial}{\partial k_{\lambda}}[k^{2}g^{\mu \nu}
-(1+ \xi^{-1})k^{\mu}k^{\nu} -
\Pi^{H \mu \nu}(k)] \frac{\partial I_{\mu \nu}^{ab}}{\partial x^{\lambda}}=
\]
\begin{equation}
= i \int \, dk^{\prime} \{ f^{bcf} f^{fde} f^{ac^{\prime}g}
f^{g d^{\prime}e^{\prime}} \,^{\ast}{\cal D}_{\mu \mu^{\prime}}^{\dagger}
(k^{\prime}) \, \tilde{\Sigma}_{k,k_{1},k_{2}, k_3}^{\mu \nu \lambda \sigma}
\tilde{\Sigma}_{k^{\prime},k_{1}^{\prime},
k_{2}^{\prime}, k_3^{\prime}}^{\dagger \mu^{\prime} \nu^{\prime}
\lambda^{\prime} \sigma^{\prime}}
\label{eq:a4}
\end{equation}
\[
\times \langle (A_{\nu^{\prime}}^{\dagger (0)c^{\prime}}(k_3^{\prime})
A_{\lambda^{\prime}}^{\dagger (0) d^{\prime}}(k_1^{\prime})
A_{\sigma^{\prime}}^{\dagger (0) e^{\prime}}(k_2^{\prime}))
(A_{\nu}^{(0)c} (k_3) A_{\lambda}^{(0) d} (k_1)
A_{\sigma}^{(0) e} (k_2)) \rangle
\]
\[
\times \delta(k - k_1 -k_2 - k_3)
\delta(k^{\prime} - k_1^{\prime} - k_2^{\prime} - k_3^{\prime})
\prod_{i=1}^{3} dk_i dk_i^{\prime} -
(a \leftrightarrow b, \, k \leftrightarrow k^{\prime}, \, compl. \,conj.) \}.
\]

\section{\bf HTL-amplitudes}
\setcounter{equation}{0}

Let us transform equation (\ref{eq:a4}) into a suitable form for
our further research. For this purpose we perform the symmetrization of the
coefficient functions in the integrand on the r.h.s. of equation (\ref{eq:a4})
over possible permutations of color and Lorentz indices,
and arguments of potentials of gauge fields within of the two groups
$(A^{(0)d}_{\lambda}(k_1)A^{(0)e}_{\sigma}(k_2)A^{(0)c}_{\nu}(k_3))$ and
$(A^{(0) \dagger d^{\prime}}_{{\lambda}^{\prime}}(k_1^{\prime})
A^{(0) \dagger e^{\prime}}_{{\sigma}^{\prime}}(k_2^{\prime})
A^{(0) \dagger c^{\prime}}_{{\nu}^{\prime}}(k_3^{\prime}))$
inside the statistical averaging angular brackets.

For example, for the first group this symmetrization leads to the expression
\[
f^{bcf}f^{fde} \, \tilde{\Sigma}^{\mu \nu \lambda \sigma}_{k,k_1,k_2,k_3}
A^{(0)d}_{\lambda}(k_1)A^{(0)e}_{\sigma}(k_2)A^{(0)c}_{\nu}(k_3) =
\frac{1}{3!} \{
f^{bcf}f^{fde} ( \tilde{\Sigma}^{\mu \nu \lambda \sigma}_{k,k_1,k_2,k_3} -
\tilde{\Sigma}^{\mu \nu \sigma \lambda}_{k,k_2,k_1,k_3})
\]
\[
+ f^{bdf}f^{fce} ( \tilde{\Sigma}^{\mu \lambda \nu \sigma}_{k,k_3,k_2,k_1} -
\tilde{\Sigma}^{\mu \lambda \sigma \nu}_{k,k_2,k_3,k_1}) +
f^{bef}f^{fdc} ( \tilde{\Sigma}^{\mu \sigma \lambda \nu}_{k,k_1,k_3,k_2} -
\tilde{\Sigma}^{\mu \sigma \nu \lambda}_{k,k_3,k_1,k_2}) \}
\]
\[
\times A^{(0)d}_{\lambda}(k_1)A^{(0)e}_{\sigma}(k_2)A^{(0)c}_{\nu}(k_3)
\]
\begin{equation}
= \frac{1}{3!} \{
[( \tilde{\Sigma}^{\mu \nu \lambda \sigma}_{k,k_1,k_2,k_3} -
\tilde{\Sigma}^{\mu \nu \sigma \lambda}_{k,k_2,k_1,k_3}) +
(\tilde{\Sigma}^{\mu \sigma \lambda \nu}_{k,k_1,k_3,k_2} -
\tilde{\Sigma}^{\mu \sigma \nu \lambda}_{k,k_3,k_1,k_2})] f^{bcf}f^{fde}
\label{eq:q5}
\end{equation}
\[
+ [( \tilde{\Sigma}^{\mu \lambda \nu \sigma}_{k,k_3,k_2,k_1} -
\tilde{\Sigma}^{\mu \lambda \sigma \nu}_{k,k_2,k_3,k_1}) -
( \tilde{\Sigma}^{\mu \sigma \lambda \nu}_{k,k_1,k_3,k_2} -
\tilde{\Sigma}^{\mu \sigma \nu \lambda}_{k,k_3,k_1,k_2})] f^{bdf}f^{fce} \}
A^{(0)d}_{\lambda}(k_1)A^{(0)e}_{\sigma}(k_2)A^{(0)c}_{\nu}(k_3).
\]
In writing the last equality, we have used the relation between the structure
constants
\[
f^{bef}f^{fdc} = f^{bcf}f^{fde} - f^{bdf}f^{fce}.
\]
In a similar way we transform the coefficient
$f^{ac^{\prime}g}f^{gd^{\prime}e^{\prime}}
\tilde{\Sigma}^{\dagger \mu^{\prime} \nu^{\prime} \lambda^{\prime} \sigma^{\prime}}_{
k^{\prime},k^{\prime}_1,k^{\prime}_2,k^{\prime}_3}$
in front of the second group of potentials of gauge fields.

The expression (\ref{eq:q5}) is convenient because it enables us to rewrite the
functions preceding the correlators on the r.h.s. of the kinetic equation (\ref{eq:a4})
in terms of HTL-amplitudes [9, 27, 28]. Actually,
by the definition of the $\tilde{\Sigma}$-function (\ref{eq:o3}) the coefficient
in front of $f^{bcf} f^{fde}$ equals
\[
\Sigma_{k,k_{1},k_{2},k_{3}}^{\mu \nu \lambda \sigma} -
\Sigma_{k,k_{2},k_{1},k_{3}}^{\mu \nu \sigma \lambda} -
\,^{\ast}{\cal D}_{\rho \alpha}(k_1 + k_2)
(S_{k,k_{3},k_{1} + k_{2}}^{\mu \nu \rho} -
S_{k,k_{1} + k_{2},k_{3}}^{\mu \rho \nu})
(S_{k_{1} + k_{2},k_{1},k_{2}}^{\alpha \lambda \sigma} -
S_{k_{1} + k_{2},k_{2},k_{1}}^{\alpha \sigma \lambda})
\]
\begin{equation}
+ ( \nu \leftrightarrow \sigma, \, k_2 \leftrightarrow k_3).
\label{eq:w5}
\end{equation}
Furthermore, we use initial definitions (\ref{eq:z2}), (\ref{eq:h2}) and
(\ref{eq:k2}).
We present the integration measure ${\rm d}^4p$ as
${\rm d} p^0 \vert{\bf p} \vert^2
{\rm d} \vert {\bf p} \vert {\rm d} \Omega$, where ${\rm d} \Omega$
is the angular measure.
Using the definition of the equilibrium distributions (\ref{eq:t2})
(for $\mu=0$) and taking into account
\[
N_c \int \limits_{- \infty}^{+ \infty} \vert {\bf p} \vert^2 \,
{\rm d} \vert {\bf p} \vert
\int \limits_{- \infty}^{+ \infty} p_0 {\rm d} p_0
\frac{d f^0(p_0)}{d p_0} = - \frac{3}{4 \pi}
\left( \frac{\omega_{pl}}{g} \right)^2,
\]
we perform the integral over ${\rm d} p_0$
and the radial integral over ${\rm d} \vert {\bf p} \vert$ in the expressions
for $S^{(II)}$-function (\ref{eq:h2}) and $\Sigma^{(II)}$-function (\ref{eq:k2}).

This enables us to present the expression (\ref{eq:w5}) in the following form
\[
-g^2 \{ \,^{\ast} \Gamma^{\mu \sigma \lambda \nu}(k,-k_2,-k_1,-k_3) -
\,^{\ast}{\cal D}_{\rho \alpha}(k_1 +k_2) \,^{\ast} \Gamma^{\mu \nu \rho}
(k,-k_3,-k_1 - k_2) \,^{\ast} \Gamma^{\alpha \lambda \sigma}(k_1 + k_2,-k_1,-k_2)
\]
\[
- \,^{\ast}{\cal D}_{\rho \alpha}(k_1 +k_3) \,^{\ast} \Gamma^{\mu \sigma \rho}(k,-k_2,
-k_1 - k_3)
\,^{\ast} \Gamma^{\alpha \lambda \nu}(k_1 + k_3,-k_1,-k_3) \}
\]
\begin{equation}
\equiv -g^2 \,^{\ast} \tilde{\Gamma}^{\mu \sigma \lambda \nu}(k,-k_2,-k_1,-k_3),
\label{eq:e5}
\end{equation}
where
\begin{equation}
\,^{\ast} \Gamma^{\mu \nu \lambda \sigma}(k,k_1,k_2,k_3) \equiv
\Gamma^{\mu \lambda \nu \sigma} + \delta \Gamma^{\mu \nu \lambda
\sigma}(k,k_3,k_1,k_2)
\label{eq:r5}
\end{equation}
is the effective four-gluon vertex, which represents a sum of a bare
four-gluon vertex
\[
\Gamma^{\mu \nu \lambda \sigma} =
2g^{\mu \nu}g^{\lambda \sigma} - g^{\mu \lambda}g^{\nu \sigma} -
g^{\mu \sigma}g^{\nu \lambda},
\]
and a corresponding HTL-correction
\[
\delta \Gamma^{\mu \nu \lambda \sigma}(k, k_1, k_2,k_3)=
m_{D}^2 \int \, \frac{{\rm d} \Omega}{4 \pi} \,
\frac{v^{\mu}v^{\nu}v^{\lambda}v^{\sigma}}{vk + i \epsilon} \, \Big[
\frac{1}{v(k + k_1) + i \epsilon} \Big( \frac{\omega_{2}}{vk_2 - i \epsilon}
\]
\[
- \frac{\omega_3}{vk_3 - i \epsilon} \Big)
- \frac{1}{v(k + k_3) + i \epsilon} \Big( \frac{\omega_{1}}{vk_1 - i \epsilon} -
\frac{\omega_2}{vk_2 - i \epsilon} \Big) \Big] \; , \; (v^{\mu} =
(1,{\bf {\bf v}}), \; v^2 = 0);
\]
\begin{equation}
\,^{\ast} \Gamma^{\mu \nu \rho}(k,k_1,k_2) \equiv
\Gamma^{\mu \nu \rho}(k,k_1,k_2) +
\delta \Gamma^{\mu \nu \rho}(k,k_1,k_2)
\label{eq:t5}
\end{equation}
is the effective three-gluon vertex. It also represents a sum of the bare
three-gluon vertex
\begin{equation}
\Gamma^{\mu \nu \rho}(k,k_1,k_2) =
g^{\mu \nu} (k - k_1)^{\rho} + g^{\nu \rho} (k_1 - k_2)^{\mu}+
g^{\mu \rho} (k_2 - k)^{\nu}
\label{eq:y5}
\end{equation}
and corresponding HTL-correction
\begin{equation}
\delta \Gamma^{\mu \nu \rho}(k,k_1,k_2) =
m_{D}^2 \int \, \frac{{\rm d} \Omega}{4 \pi} \,
\frac{v^{\mu}v^{\nu}v^{\rho}}{vk + i \epsilon} \,
\Big( \frac{\omega_2}{vk_2 - i \epsilon} -
\frac{\omega_1}{vk_1 - i \epsilon} \Big),
\label{eq:u5}
\end{equation}
$m_{D}^2 = 3 \, \omega^2_{pl}$ is the Debye screening mass.

The polarization tensor (\ref{eq:f2}) in this notation takes the form
\[
\Pi^{\mu \nu}(k) = m_{D}^2
\left( g^{\mu 0}g^{\nu 0} - \omega \int \, \frac{{\rm d} \Omega}{4 \pi}
\frac{v^{\mu}v^{\nu}}{vk + i \epsilon} \right).
\]

In a similar way, the coefficient in front of the product of structure constants
$f^{bdf}f^{fce}$ in (\ref{eq:q5}) may be presented as
\[
-g^2 \{ \,^{\ast} \Gamma^{\mu \lambda \nu \sigma}(k,-k_1,-k_3,-k_2) -
\,^{\ast}{\cal D}_{\rho \alpha}(k_2 +k_3) \,^{\ast} \Gamma^{\mu \lambda \rho}
(k,-k_1,-k_2 - k_3) \,^{\ast} \Gamma^{\alpha \nu \sigma}(k_2 + k_3,-k_3,-k_2)
\]
\[
+ \,^{\ast}{\cal D}_{\rho \alpha}(k_1 +k_3) \,^{\ast} \Gamma^{\mu \sigma \rho}(k,-k_2,
-k_1 - k_3)
\,^{\ast} \Gamma^{\alpha \lambda \nu}(k_1 + k_3,-k_1,-k_3) \}
\]
\begin{equation}
\equiv -g^2 \,^{\ast} \tilde{\Gamma}^{\mu \lambda \nu \sigma}(k,-k_1,-k_3,-k_2).
\label{eq:i5}
\end{equation}

Performing transformation of the coefficient preceding
$(A^{(0) \dagger d^{\prime}}_{{\lambda}^{\prime}}(k_1^{\prime})
A^{(0) \dagger e^{\prime}}_{{\sigma}^{\prime}}(k_2^{\prime})
A^{(0) \dagger c^{\prime}}_{{\nu}^{\prime}}(k_3^{\prime}))$
in a similar manner, we can cast the equation (\ref{eq:a4}) in the following form
\begin{equation}
\frac{\partial}{\partial k_{\lambda}} [k^2 g^{\mu \nu} -
(1 + \xi^{-1}) k^{\mu} k^{\nu} - \Pi^{H \mu \nu} (k) ]
\frac{\partial I_{\mu \nu}^{a b}}{\partial x^{\lambda}}
\label{eq:o5}
\end{equation}
\[
= i \frac{g^{4}}{(3!)^2} \int \, dk^{\prime} \,
[ \{ f^{bcf} f^{fde} \,^{\ast} \tilde{\Gamma}^{\mu \sigma \lambda \nu}
(k, - k_2, - k_1, - k_3) +
f^{bdf} f^{fce} \,^{\ast} \tilde{\Gamma}^{\mu \lambda \nu \sigma}
(k, - k_1, - k_3, - k_2) \}
\]
\[
\times \,^{\ast}{\cal D}_{\mu \mu^{\prime}}^{\dagger}(k^{\prime})
\{ f^{ac^{\prime}g} f^{gd^{\prime}e^{\prime}} \,^{\ast}
\tilde{\Gamma}^{\dagger \mu^{\prime} \sigma^{\prime} \lambda^{\prime} \nu^{\prime}}
(k^{\prime}, - k_2^{\prime}, - k_1^{\prime}, - k_3^{\prime})
+ f^{a d^{\prime} g} f^{g c^{\prime} e^{\prime}} \,^{\ast}
\tilde{\Gamma}^{\dagger \mu^{\prime} \lambda^{\prime} \nu^{\prime} \sigma^{\prime}}
(k, - k_1^{\prime}, - k_3^{\prime}, - k_2^{\prime}) \}
\]
\[
\times \langle A_{\nu^{\prime}}^{\dagger (0) c^{\prime}} (k_3^{\prime})
A_{\lambda^{\prime}}^{\dagger (0) d^{\prime}} (k_1^{\prime})
A_{\sigma^{\prime}}^{\dagger (0) e^{\prime}} (k_2^{\prime})
A_{\nu}^{(0) c} (k_3) A_{\lambda}^{(0) d} (k_1) A_{\sigma}^{(0) e} (k_2) \rangle
\]
\[
\times \delta(k - k_1 - k_2 - k_3)
\delta(k^{\prime} - k_1^{\prime} - k_2^{\prime} - k_3^{\prime})
\prod_{i=1}^{3} d k_i d k_{i}^{\prime} - (a \leftrightarrow b, \,
k \leftrightarrow k^{\prime}, \, compl. conj.)].
\]
The r.h.s. of equation (\ref{eq:o5})
has a non-trivial color structure that actually is well represented by
non-trivial color
structure of the initial dynamical equation (\ref{eq:r2}).
As will be shown bellow, this leads to a qualitative distinction between
the elastic scattering probability $w({\bf k}, {\bf k}_1; {\bf k}_2, {\bf k}_3)$
of colorless plasmons in a hot QCD plasma and a similar one of plasmons in a hot QED
plasma [24, 30].

At the end of this Section we present the identities analogous to
the effective Ward one in hot gauge theories [27, 28, 9]. It can be
shown that the following equalities hold
\begin{equation}
k_{\mu} \,^{\ast} \Gamma^{\mu \nu \lambda \sigma}(k,k_1,k_2,k_3) =
\,^{\ast} \Gamma^{\nu \lambda \sigma}(k_1,k_2,k+k_3) -
\,^{\ast} \Gamma^{\nu \lambda \sigma}(k+k_1,k_2,k_3),
\label{eq:p5}
\end{equation}
\[
k_{1 \nu} \,^{\ast} \Gamma^{\mu \nu \lambda \sigma}(k,k_1,k_2,k_3) =
\,^{\ast} \Gamma^{\mu \lambda \sigma}(k+k_1,k_2,k_3) -
\,^{\ast} \Gamma^{\mu \lambda \sigma}(k,k_1+k_2,k_3)
\]
(similar contractions with $k_{2 \lambda}, k_{3 \sigma}$),
\begin{equation}
k_{\mu} \,^{\ast} \Gamma^{\mu \nu \rho}(k,k_1,k_2) =
\,^{\ast}{\cal D}^{-1 \, \nu \rho}(-k_1) - \,^{\ast}{\cal D}^{-1 \, \nu \rho}(-k_2)
\label{eq:a5}
\end{equation}
(similar contractions with $k_{1 \nu}, k_{2 \rho}$).
Here, $\,^{\ast}{\cal D}^{-1 \, \mu \nu}(k) = P^{\mu \nu }(k)
\,^{\ast} \Delta^{-1 \, t}(k) +
Q^{\mu \nu}(k) \,^{\ast} \Delta^{-1 \, l}(k)$ is the inverse propagator
without the gauge fixing term.

\section{\bf The kinetic equation for plasmons}
\setcounter{equation}{0}

We are now in a position to explicitly compute the probability of plasmon-plasmon
scattering. First of all, we make the correlation decoupling of the
sixth-order correlators on the r.h.s. of equation (\ref{eq:o5}) in terms of
the pair ones by the rule (\ref{eq:i4}). After cumbersome calculations,
the r.h.s. of equation (\ref{eq:o5}) can be written as
\[
g^4 \,{\rm Im}\, (\,^{\ast}{\cal D}_{\mu \mu^{\prime}} (k)) \int \, \{
f^{bcf} f^{fde} \,^{\ast} \tilde{\Gamma}^{\mu \sigma \lambda \nu}
(k, - k_2, - k_1, - k_3) +
f^{bd f} f^{fce} \,^{\ast} \tilde{\Gamma}^{\mu \lambda\nu \sigma}
(k, - k_1, - k_3, - k_2) \}
\]
\begin{equation}
\times \{ f^{acg} f^{gde} \,^{\ast} \tilde{\Gamma}^{\dagger \mu^{\prime} \sigma^{\prime}
\lambda^{\prime} \nu^{\prime}}
(k, - k_2, - k_1, - k_3)
+ f^{adg} f^{gce} \,^{\ast} \tilde{\Gamma}^{\dagger \mu^{\prime} \lambda^{\prime}
\nu^{\prime} \sigma^{\prime}}
(k, - k_1, - k_3, - k_2) \}
\label{eq:q6}
\end{equation}
\[
\times I_{\nu \nu^{\prime}} (k_3)
I_{\lambda \lambda^{\prime}} (k_1)
I_{\sigma \sigma^{\prime}} (k_2)
\delta(k - k_1 - k_2 - k_3)  dk_1 dk_2 dk_3 .
\]

In deriving of (\ref{eq:q6}) we have used two relations, which satisfy
$\,^{\ast}\tilde{\Gamma}^{\mu \nu \lambda \sigma} (k, k_1, k_2, k_3)$
\[
\,^{\ast} \tilde{\Gamma}^{\mu \nu \lambda \sigma} (k, k_1, k_2, k_3) +
\,^{\ast} \tilde{\Gamma}^{\mu \lambda \nu \sigma} (k, k_2, k_1, k_3) +
\,^{\ast} \tilde{\Gamma}^{\mu \nu \sigma \lambda} (k, k_1, k_3, k_2) = 0,
\]
\begin{equation}
\,^{\ast} \tilde{\Gamma}^{\mu \nu \lambda \sigma} (k, k_1, k_2, k_3) =
\,^{\ast} \tilde{\Gamma}^{\mu \sigma \lambda \nu} (k, k_3, k_2, k_1) .
\label{eq:w6}
\end{equation}
Their correctness may be verified by a direct calculation using known properties
of HTL-amplitudes [27, 28], entering into the definition of
$\,^{\ast} \tilde{\Gamma}^{\mu \nu \lambda \sigma} (k, k_1, k_2, k_3)$
(\ref{eq:e5}).
The first of the relations in (\ref{eq:w6}) supplements, in the some sense,
a similar relation for the HTL-correction $\delta \Gamma_4$ to the bare
four-gluon vertex,
originally proposed by Frenkel and Taylor [28]. The second one
is a property of invariance of the function
$\,^{\ast} \tilde{\Gamma}$, when the momenta order is reversed.
Notice that order of the space-time indices in equation (\ref{eq:w6}) is important.

Furthermore, if we restrict our consideration to the study of plasmons scattering by
plasmons, then in the spectral decomposition of $I_{\mu \mu^{\prime}}$
into the orthonormal projectors
$P_{\mu \mu^{\prime}}$ and $Q_{\mu \mu^{\prime}}$, it is necessary to keep only
the longitudinal part, i.e. set
\begin{equation}
I_{\nu \nu^{\prime}} (k_3) = Q_{\nu \nu^{\prime}} (k_3) I^l (k_3) \; , \;
I_{\lambda \lambda^{\prime}} (k_1) = Q_{\lambda \lambda^{\prime}}
(k_1) I^l (k_1) \; , \;
I_{\sigma \sigma^{\prime}} (k_2) = Q_{\sigma \sigma^{\prime}} (k_2) I^l (k_2).
\label{eq:e6}
\end{equation}
In the propagator $\,^{\ast}{\cal D}_{\mu \mu^{\prime}} (k)$  we also keep only
the longitudinal part $- Q_{\mu \mu^{\prime}} (k) \,^{\ast} \Delta^l (k)$.
To determine the imaginary part of $\,^{\ast} \Delta^l (k) = 1/(k^2 - \Pi^l (k))$
we use the approximation (see e.g., Pustovalov and Silin in [22])
\[
\frac{1}{k^2 - \Pi^l (k)} \approx \frac{{\rm P}}{{\rm Re} (k^2 - \Pi^l (k))}
- i \pi \, sign \, ({\rm Im} \, (k^2 - \Pi^l (k)) \delta
({\rm Re} \, (k^2 - \Pi^l(k)) ) =
\]
\begin{equation}
= \frac{{\rm P}}{k^2 {\rm Re} \, \varepsilon^{l} (k)} - \frac{i \pi}{k^2}
sign \, \omega \, \delta ({\rm Re} \, \varepsilon^l (k) ) ,
\label{eq:r6}
\end{equation}
which holds for a small ${\rm Im} \, \varepsilon^l (k)$.
Here, we consider that because of (\ref{eq:e1}), $sign \,({\rm Im} \, \varepsilon^l
(k) ) = sign \, \omega$. The symbol ${\rm P}$ denotes a principal value.
We perceive the
$\delta$-function of the real part of the longitudinal permeability, which appears
in (\ref{eq:r6}) in the ordinary sense (\ref{eq:e4}).
Substituting (\ref{eq:e4}) into (\ref{eq:r6}) and choosing
$\omega = \omega_{\bf k}^l > 0$ for definiteness, we obtain the required relation
\begin{equation}
{\rm Im} \,^{\ast}{\cal D}_{\mu \mu^{\prime}} (k) \approx \frac{\pi}{k^2}
Q_{\mu \mu^{\prime}} (k)
\bigg( \frac{\partial {\rm Re} \, \varepsilon^l (k)}{\partial \omega}
\bigg)^{-1}_{\omega = \omega_{\bf k}^l} \delta ( \omega - \omega_{\bf k}^l) .
\label{eq:t6}
\end{equation}

With allowing (\ref{eq:e6}), (\ref{eq:t6}) and performing the color algebra
with the following identities
\[
f^{bcf} f^{fde} f^{acg} f^{gde} = N_c^2 \delta^{ab} ,
\]
\begin{equation}
f^{bcf} f^{fde} f^{adg} f^{gce} = \frac{1}{2} N_c^2 \delta^{ab} ,
\label{eq:y6}
\end{equation}
we rewrite (\ref{eq:q6}) in the form
\[
\delta^{ab} \pi g^4 N_c^2 \frac{1}{k^2}
\Big( \frac{\partial {\rm Re} \, \varepsilon^l (k)}{\partial \omega}
\Big)^{- 1}_{\omega= \omega_{\bf k}^l}
\delta(\omega - \omega_{\bf k}^l) \int
\{
\vert \,^{\ast} \tilde{\Gamma} (k, -k_2, -k_1,-k_3) \vert^2 +
\vert \,^{\ast} \tilde{\Gamma} (k, -k_1, -k_3,-k_2) \vert^2
\]
\begin{equation}
+ {\rm Re} \, (\,^{\ast} \tilde{\Gamma} (k, -k_2, -k_1,-k_3)
\,^{\ast} \tilde{\Gamma}^{\dagger} (k, -k_1, -k_3,-k_2)) \} \,
\frac{1}{\bar{u}^2(k)
\bar{u}^2(k_1)
\bar{u}^2(k_2)
\bar{u}^2(k_3)}
\label{eq:u6}
\end{equation}
\[
\times I^l (k_1)
I^l (k_2)
I^l (k_3)
\delta(k - k_1 - k_2 - k_3) dk_1 dk_2 dk_3 .
\]
Here, we denote
\begin{equation}
\,^{\ast} \tilde{\Gamma} (k, k_1, k_2, k_3) \equiv
\,^{\ast} \tilde{\Gamma}^{\mu \nu \lambda \sigma} (k, k_1, k_2, k_3)
\bar{u}_{\mu} (k)
\bar{u}_{\nu} (k_1)
\bar{u}_{\lambda} (k_2)
\bar{u}_{\sigma} (k_3).
\label{eq:i6}
\end{equation}
By virtue of the fact, that the spectral intensity functions
$I^l (k_i), i = 1, 2, 3$, entering into (\ref{eq:u6}), satisfy the equations
${\rm Re} \, \varepsilon^l (k_i) I^l (k_i) = 0$ [10], they have the following
structure
\begin{equation}
I^l (k_i) = I_{{\bf k}_i}^l \delta (\omega_i - \omega_{{\bf k}_i}^l)
+ I_{-{\bf k}_i}^l \delta (\omega_i + \omega_{{\bf k}_i}^l)  \; , \;
i = 1, 2, 3 ,
\label{eq:o6}
\end{equation}
where $I_{{\bf k}_i}^l$ are certain functions of the wavevectors ${\bf k}_i$.

We substitute (\ref{eq:o6}) into (\ref{eq:u6}) and perform the integration
over the frequency ${\rm d} \omega_i$ with the help of the $\delta$-functions.
The function $I_k^l$ in the l.h.s. of the kinetic equation (\ref{eq:o5})
also has a structure of form (\ref{eq:o6}). Here, we keep only a positive
branch $\omega > 0$ in agreement with (\ref{eq:t6}).
Furthermore performing integration (\ref{eq:o5}) over ${\rm d} \omega$,
where the r.h.s. of (\ref{eq:o5}) has the form (\ref{eq:u6}), as result
we obtain
\begin{equation}
k^2 \Big( \frac{\partial {\rm Re} \, \varepsilon^l (k)}{\partial \omega}
\Big)_{\omega = \omega_{\bf k}^l} \frac{\partial I_{\bf k}^l}
{\partial t} + (- k^2)
\Big( \frac{\partial {\rm Re} \, \varepsilon^l (k)}{\partial {\bf k}}
\Big)_{\omega = \omega_{\bf k}^l} \frac{\partial I_{\bf k}^l}
{\partial {\bf x}}
\label{eq:p6}
\end{equation}
\[
= \pi g^4 N_c^2 \frac{1}{k^2} \Big( \frac{\partial {\rm Re} \,
\varepsilon^l (k)}{\partial \omega} \Big)_{\omega = \omega_{\bf k}^l}^{- 1}
\int \, [ Q (k, k_1, k_2, k_3) \delta(k + k_1 + k_2 + k_3)
\]
\[
+ Q (k,- k_1,- k_2,- k_3) \delta(k - k_1 - k_2 - k_3)
+ Q (k, k_1, k_2,- k_3) \delta(k + k_1 + k_2 - k_3)
\]
\[
+ Q (k, k_1,- k_2, k_3) \delta(k + k_1 - k_2 + k_3) +
Q(k,- k_1, k_2, k_3) \delta(k - k_1 +  k_2 + k_3)
\]
\[
+ Q (k, k_1,- k_2,- k_3) \delta(k + k_1 - k_2 - k_3) +
Q (k,- k_1, k_2,- k_3) \delta(k - k_1 + k_2 - k_3)
\]
\[
+ Q (k,- k_1,- k_2, k_3) \delta(k - k_1 - k_2 + k_3) ]_{on-shell}
I_{{\bf k}_1}^l
I_{{\bf k}_2}^l
I_{{\bf k}_3}^l
\, {\rm d} {\bf k}_1
\, {\rm d} {\bf k}_2
\, {\rm d} {\bf k}_3 ,
\]
where we have defined
\begin{equation}
Q (k, k_1, k_2, k_3) \equiv
\frac{1}{\bar{u}^2(k)
\bar{u}^2(k_1)
\bar{u}^2(k_2)
\bar{u}^2(k_3)}
\label{eq:a6}
\end{equation}
\[
\times \{
\vert \,^{\ast} \tilde{\Gamma} (k, k_2, k_1, k_3) \vert^2 +
\vert \,^{\ast} \tilde{\Gamma} (k, k_1, k_3, k_2) \vert^2 +
{\rm Re} \, (\,^{\ast} \tilde{\Gamma} (k, k_2, k_1, k_3)
\,^{\ast} \tilde{\Gamma}^{\dagger} (k, k_1, k_3, k_2) ) \}.
\]
In equation (\ref{eq:p6}) the $x$-dependence of $I^l_{\bf k}, \, I^l_{{\bf k}_1}$,
etc is understood, although not written explicitly.

The first term inside the square brackets on the r.h.s. of (\ref{eq:p6}) describes the
process of simultaneous fusion or emission of four plasmons in the plasma.
Considering that the $\delta$-function has no support on the plasmon mass shell,
its contribution to the kinetic equation
vanishes. The second, third and fourth terms describe the decay of one plasmon
into three or the fusion of three plasmons into one. As was mentioned in
section 2, these processes are forbidden by the conservation law
(\ref{eq:v1}). Therefore these terms also vanish. The remaining
terms describe the scattering of two plasmons off two plasmons.
Three terms imply the existence of three possible channels of a given process, which
change the plasmon numbers density $N_{\bf k}^l$
\[
{\rm g}^{\ast} + {\rm g}_1^{\ast} \rightleftharpoons
{\rm g}_2^{\ast} + {\rm g}_3^{\ast}, \;
{\rm g}^{\ast} + {\rm g}_2^{\ast} \rightleftharpoons
{\rm g}_1^{\ast} + {\rm g}_3^{\ast}, \;
{\rm g}^{\ast} + {\rm g}_3^{\ast} \rightleftharpoons
{\rm g}_1^{\ast} + {\rm g}_2^{\ast} .
\]
If we perform replacements ${\bf k}_1 \leftrightarrow {\bf k}_2$ in the next
to last term on the r.h.s. of (\ref{eq:p6}), and
$ {\bf k}_1 \leftrightarrow {\bf k}_3$ in the last term,
then the r.h.s. of (\ref{eq:p6}) can be presented as
\[
 \pi g^4 N_c^2 \frac{1}{k^2} \Big( \frac{\partial {\rm Re} \,
\varepsilon^l (k)}{\partial \omega} \Big)_{\omega = \omega_{\bf k}^l}^{- 1}
\int \, [ Q (k, k_1,- k_2,- k_3)
+ Q (k,- k_2, k_1,- k_3) +
Q (k,- k_3,- k_2, k_1 )]_{on-shell}
\]
\[
\times \delta(\omega_{\bf k}^l + \omega_{{\bf k}_1}^l - \omega_{{\bf k}_2}^l -
\omega_{{\bf k}_3}^l)
\delta({\bf k} + {\bf k_1} - {\bf k_2} - {\bf k_3})
I_{{\bf k}_1}^l
I_{{\bf k}_2}^l
I_{{\bf k}_3}^l
{\rm d} {\bf k}_1
{\rm d} {\bf k}_2
{\rm d} {\bf k}_3.
\]
By using the definition of kernel $Q$ (\ref{eq:a6}) and the
properties (\ref{eq:w6}), it can be shown, that
\[
Q(k, -k_2, k_1, -k_3) +
Q(k, -k_3, - k_2, k_1) =
2 Q(k, k_1, -k_2, -k_3).
\]

Taking into account the last relation, it goes over from the function
$I_{\bf k}^l$ to the number density of plasmons
\[
 N_{\bf k}^l  = - \Big( k^2 \frac{\partial {\rm Re} \,
\varepsilon^l (k)}{\partial \omega} \Big)_{\omega = \omega_{\bf k}^l}
I_{\bf k}^l
\]
and recovering the complete form for finite values $N^l_{\bf k}$, we obtain the
required Boltzmann equation (\ref{eq:n1}), where the probability of
plasmon-plasmon scattering is defined by the following expression
\begin{equation}
w ({\bf k}, {\bf k}_1; {\bf k}_2, {\bf k}_3) =
3 \pi (2 \pi)^5 g^4 N_c^2  \, \bigg[ \frac{1}{k^2 k_1^2 k_2^2 k_3^2}
\Big( \frac{\partial {\rm Re} \, \varepsilon^l (k)}{\partial \omega}
\Big)^{- 1} \Big( \frac{\partial {\rm Re} \, \varepsilon^l (k_1)}
{\partial \omega_1} \Big)^{- 1}
\label{eq:f6}
\end{equation}
\[
\times \Big( \frac{\partial {\rm Re} \, \varepsilon^l (k_2)}
{\partial \omega_2} \Big)^{- 1} \Big( \frac{\partial {\rm Re} \,
\varepsilon^l (k_3)}{\partial \omega_3} \Big)^{- 1}
\{ \vert \,^{\ast} \tilde{\Gamma} (k, - k_2, k_1, - k_3) \vert^2 +
\vert \,^{\ast} \tilde{\Gamma} (k, k_1, - k_3, - k_2) \vert^2
\]
\[
+ {\rm Re} \,( \,^{\ast} \tilde{\Gamma}(k, - k_2, k_1, - k_3)
 ^{\ast} \tilde{\Gamma}^{\dagger}(k, k_1, - k_3, - k_2)) \}
\frac{1}{\bar{u}^2(k)
\bar{u}^2(k_1) \bar{u}^2(k_2) \bar{u}^2(k_3)} \bigg]_{on-shell}.
\]

The result (\ref{eq:f6}) is rather unexpected.
As we can see from the expression (\ref{eq:f6}), this probability does not reduced
to the squared module of one scalar function, as this occurs in the Abelian
plasma. Here, the scattering probability is
defined by the squared module of two independent scalar functions and their
interference\footnote{It is easy to see from the expression (\ref{eq:f6}),
that the function ${\it w}({\bf k}, {\bf k}_1; {\bf k}_2, {\bf k}_3)$
is positively definite.}, in spite of the fact that in this paper we
restrict our consideration to a study of the nonlinear interaction of only colorless
excitations. This radically distinguishes the Boltzmann equation
(\ref{eq:n1}), describing the effects of the collisions among colorless
soft excitations from the corresponding Boltzmann
equation including the effects of the collisions among colorless hard
excitations [34, 18, 19]. In the last case, the Boltzmann
equation, corrected to color factors, fully coincides with the corresponding one
in the Abelian plasma. This is a point which we find difficult to interpret and
therefore additional analysis of this problem is required (see, also discussion
in conclusion).

The scattering probability can be written in the form, which is manifestly
symmetric under permutations of the external momenta ${\bf k}, {\bf k}_1,
{\bf k}_2$ and ${\bf k}_3$
\[
w({\bf k}, {\bf k}_1; {\bf k}_2, {\bf k}_3) =
 \pi(2 \pi)^5 g^4 N_c^2  \bigg[ \frac{1}{k^2 k_1^2 k_2^2 k_3^2}
\Big( \frac{\partial {\rm Re} \, \varepsilon^l (k)}{\partial \omega} \Big)^{- 1}
\Big( \frac{\partial {\rm Re} \, \varepsilon^l (k_1)}{\partial \omega_1} \Big)^{- 1}
\Big( \frac{\partial {\rm Re} \, \varepsilon^l (k_2)}{\partial \omega_2} \Big)^{- 1}
\]
\[
\times \Big( \frac{\partial {\rm Re} \, \varepsilon^l (k_3)}{\partial \omega_3}
\Big)^{- 1} \{ \vert \,^{\ast} \tilde{\Gamma} (k, - k_2, k_1, - k_3) \vert^2 +
\vert \,^{\ast} \tilde{\Gamma} (k, k_1, - k_3, - k_2) \vert^2 +
\vert \,^{\ast} \tilde{\Gamma} (k,  k_1,- k_2, - k_3) \vert^2 +
\]
\[
{\rm Re} \,( \,^{\ast} \tilde{\Gamma}(k, - k_2, k_1, - k_3)
\,^{\ast} \tilde{\Gamma}^{\dagger}(k, k_1, - k_3, - k_2) )
+ {\rm Re} \, (\,^{\ast} \tilde{\Gamma}(k, - k_2, k_1, - k_3)
\,^{\ast} \tilde{\Gamma}^{\dagger}(k, k_1, - k_2, - k_3))
\]
\[
+ {\rm Re} \,( \,^{\ast} \tilde{\Gamma}(k, k_1,- k_3, - k_2)
\,^{\ast} \tilde{\Gamma}^{\dagger}(k, k_1, - k_2, - k_3) ) \}
\frac{1}{\bar{u}^2(k) \bar{u}^2(k_1) \bar{u}^2(k_2) \bar{u}^2(k_3)} \bigg]_{on-shell}.
\]
This expression is suitable for checking the symmetry conditions (\ref{eq:m1}),
which is imposed on the plasmon-plasmon scattering probability.

Recall that the function $\,^{\ast} \tilde{\Gamma}$, which appears in the expression for
probability (\ref{eq:f6}), is defined by expressions (\ref{eq:i6}) and
(\ref{eq:e5}). As we have shown in [10], the expression
(\ref{eq:i6}) (exactly, its part, independent on a gauge-parameter) is gauge
invariant at least in a class of covariant and temporal gauges.
With regard to the reasoning in section 4
this automatically leads to gauge invariance of the kinetic equation
(\ref{eq:n1}). The problem of the dependence of the
plasmon-plasmon scattering probability (\ref{eq:f6}) on a gauge parameter
coming from the gauge fixing term in the gluon propagator, is discussed below.

Let us separate the terms with a gauge parameter from the expression
$\,^{\ast} \tilde{\Gamma} (k, -k_2, k_1, -k_3)$.
By using the Ward identities (\ref{eq:p5}) and (\ref{eq:a5}), we have
\[
\xi \{D_{\rho \alpha}(-k_1 + k_2) \Delta^{0}(-k_1 + k_2)
\,^{\ast}\Gamma^{\mu \nu \rho}(k, -k_3,k_1 - k_2)
\,^{\ast}\Gamma^{\alpha \lambda \sigma}(-k_1+k_2,k_1,-k_2)
\]
\[
+ D_{\rho \alpha}(-k_1 + k_3) \Delta^{0}(-k_1 + k_3)
\,^{\ast}\Gamma^{\mu \sigma \rho}(k, -k_2, k_1 - k_3)
\,^{\ast}\Gamma^{\alpha \lambda \nu}(-k_1+k_3,k_1,-k_3) \}
\]
\begin{equation}
\times \bar{u}_{\mu} (k) \bar{u}_{\lambda} (k_1) \bar{u}_{\sigma} (k_2)
\bar{u}_{\nu} (k_3)
\label{eq:g6}
\end{equation}
\[
= \xi \{(\Delta^{0}(-k_1 + k_2))^2 ( \,^{\ast}{\cal D}^{-1 \, \mu \sigma}(k)
- \,^{\ast}{\cal D}^{-1 \, \mu \sigma}(k_3))
( \,^{\ast}{\cal D}^{-1 \, \lambda \nu}(k_2) - \,^{\ast}
{\cal D}^{-1 \, \lambda \nu}(-k_1))
\]
\[
+ (\Delta^{0}(-k_1 + k_3))^2 ( \,^{\ast}{\cal D}^{-1 \, \mu \nu}(k)
- \,^{\ast}{\cal D}^{-1 \, \mu \nu}(k_2))
(\,^{\ast}{\cal D}^{-1 \, \lambda \sigma}(k_3) -
\,^{\ast}{\cal D}^{-1 \, \lambda \sigma}(-k_1)) \}
\]
\[
\times \bar{u}_{\mu}(k) \bar{u}_{\lambda}(k_1)
\bar{u}_{\sigma}(k_2) \bar{u}_{\nu}(k_3).
\]
It is easily shown that expression on the r.h.s. of (\ref{eq:g6}) vanish either
because $\,^{\ast}{\cal D}^{- 1 \, \mu \nu} (k)$ is transverse, or
by the definition of the mass-shall condition, i.e.
\begin{equation}
k_{\mu} \,^{\ast}{\cal D}^{-1 \, \mu \nu}(k) = 0, \;
\,^{\ast}{\cal D}^{-1 \, 0 \nu}(k) \vert_{\omega= \omega^{l}_{\bf k}} = 0.
\label{eq:h6}
\end{equation}
If we carry out the following replacements on the r.h.s. of the expressions
(\ref{eq:i6}), (\ref{eq:e5})
\[
\bar{u}_{\mu}(k) \rightarrow \tilde{u}_{\mu}(k) \equiv
\frac{k^2}{(ku)}(k_{\mu} - u_{\mu}(ku)),
\]
and
\[
\,^{\ast}{\cal D}_{\rho \alpha}(k) \rightarrow
\,^{\ast}\tilde{\cal D}_{\rho \alpha}(k) = \,^{\ast}{\cal D}_{\rho \alpha}(k) -
\Big( \frac{\sqrt{-2k^2 {\bar{u}}^2}}{k^2(ku)}C_{\rho \alpha}(k) +
\frac{\bar{u}^2(k)}{k^2(ku)^2}D_{\rho \alpha}(k) \Big) \,^{\ast} \Delta^{l}(k)
\]
\[
- \xi D_{\rho \alpha}(k) \Delta^{0}(k) -
\xi_{0} \frac{k^2}{(ku)^2} D_{\rho \alpha}(k),
\]
where $\xi_{0}$ is a gauge parameter in the temporal gauge, then
it can be proved that a similar statement holds in the temporal gauge also.

Thus, the gauge-dependent parts of $w({\bf k},{\bf k}_1;{\bf k}_2,{\bf k}_3)$
drop out,
since they are multiplied by the mass-shell factor.
These factors are proportional to $(\omega - \omega_{\bf k}^{l})$.
However, in the quantum case Baier {\it et al} [35]
observed that a naive calculation of the gluon damping rate in a covariant gauge
appears to violate this consideration. The mass-shell factor is multiplied
by the integral involving a power infrared divergence which is cut-off
exactly on the scale $(\omega - \omega_{\bf k}^{l}) \sim g^2 T$.
This problem was careful discussed in consideration of the nonlinear Landau
damping [10], when in (\ref{eq:e5}) and consequently (\ref{eq:g6}) it is
necessary to set
\begin{equation}
k_1 = - k, \; k_2 = - k_1, \; k_3 = k_1.
\label{eq:j6}
\end{equation}
We have shown, that the gauge-dependent part of the nonlinear Landau damping
rate vanishes on a mass-shell at least for the plasmons with zero momentum.
In the case of the process of the plasmon-plasmon scattering considered here, the problem
on the dependence of $w({\bf k},{\bf k}_1;{\bf k}_2,{\bf k}_3)$
on a gauge parameter is more subtle, since instead of (\ref{eq:j6}), we have
condition: $k + k_1 = k_2 + k_3$ only. Research into this nontrivial question goes
beyond our present goal.

\section{\bf Lifetimes of plasmons}
\setcounter{equation}{0}

To calculate the lifetimes of colorless plasmons we first linearize the
Boltzmann equation (\ref{eq:n1}) (here, we assume that off-equilibrium
fluctuation is perturbative small), writing the number density of plasmons as
\[
N^l_{\bf k} = N_{eq}^l({\bf k}) + \delta N^l_{\bf k}
\]
where $N_{eq}^{l}({\bf k}) = (e^{\omega^{l}_{\bf k}/T^{\ast}} - 1)^{-1}$
is the Planck distribution function and $T^{\ast}$ is a certain constant, which can be interpreted as a
plasmon gas temperature in the statistical equilibrium state.
Then we find
\begin{equation}
\frac{\partial \delta N_{\bf k}^l}{\partial t} +
{\bf V}^l_{\bf k}
\frac{\partial \delta N_{\bf k}^l}{\partial {\bf x}} =
\label{eq:q7}
\end{equation}
\[
= \int \frac{{\rm d} {\bf k}_1}{(2 \pi)^3} \frac{{\rm d} {\bf k}_2}{(2 \pi)^3}
\frac{{\rm d} {\bf k}_3}{(2 \pi)^3} \,
(2 \pi)^4 \delta (\omega_{\bf k}^l + \omega_{{\bf k}_1}^l - \omega_{{\bf k}_2}^l -
\omega_{{\bf k}_3}^l)
\delta ({\bf k} + {\bf k}_1 - {\bf k}_2 - {\bf k}_3)
{\it w}({\bf k},{\bf k}_1;{\bf k}_2,{\bf k}_3)
\]
\[
\times \{ \delta N^l_{\bf k}[ N_{eq}^l({{\bf k}_2}) N^l_{eq}({{\bf k}_3})
( N^l_{eq}({{\bf k}_1}) + 1 ) - N_{eq}^l({{\bf k}_1})
( N_{eq}^l({{\bf k}_2}) + 1 ) ( N_{eq}^l({{\bf k}_3}) + 1 )] +
\]
\[
\delta N^l_{{\bf k}_2}[ N_{eq}^l({{\bf k}_3})
(N^l_{eq}({\bf k}) + 1)
( N^l_{eq}({{\bf k}_1}) + 1 ) - N_{eq}^l({\bf k})
N_{eq}^l({{\bf k}_1})( N_{eq}^l({{\bf k}_3}) + 1 )]
+ ({\bf k} \leftrightarrow {\bf k}_1, \, {\bf k}_2 \leftrightarrow {\bf k}_3) \}.
\]
This equation can be further simplified if we use the following
parametrization for off-equilibrium fluctuation of the occupation number
$\delta N^l_{\bf k}$ [12, 18]
\begin{equation}
\delta N^l_{\bf k} \equiv - \frac{{\rm d} N^l_{eq}({\bf k})}
{{\rm d} \omega^l_{\bf k}}{\cal W}^l_{\bf k} =
(1/T^{\ast}) N^l_{eq}({\bf k})(N^l_{eq}({\bf k}) + 1) {\cal W}^l_{\bf k}.
\label{eq:w7}
\end{equation}
Let us introduce the momentum transfers, which carries
the soft gluon exchanged in the collision of two plasmons, setting
\begin{equation}
{\bf k}_2 = {\bf k} - {\bf q}, \; {\bf k}_3 = {\bf k}_1 + {\bf q}.
\label{eq:e7}
\end{equation}
Performing the integration over ${\rm d}{\bf k}_3$, replacing the
integration over ${\rm d}{\bf k}_2$ by one with respect to
momentum transfer and taking into account (\ref{eq:w7}) and (\ref{eq:e7}),
finally we derive a linearized Boltzmann equation for ${\cal W}^l_{\bf k}$
function
\[
\frac{\partial {\cal W}^l_{\bf k}}{\partial t} +
{\bf V}^l_{\bf k}
\frac{\partial {\cal W}^l_{\bf k}}{\partial {\bf x}} =
- \int \frac{{\rm d}{\bf k}_1}{(2 \pi)^3}
\int \frac{{\rm d}{\bf q}}{(2 \pi)^3} \,
\frac{N_{eq}^l({\bf k}_1)(N_{eq}^l({\bf k}-{\bf q}) + 1)
(N_{eq}^l({\bf k}_1+{\bf q}) + 1)}
{(N_{eq}^l({\bf k}) + 1)}
\]
\begin{equation}
\times {\it w}({\bf k},{\bf k}_1;{\bf q})
\, (2 \pi) \delta(\omega^l_{\bf k} - \omega^l_{{\bf k}-{\bf q}} +
\omega^l_{{\bf k}_1} - \omega^l_{{\bf k}_1 + {\bf q}})
\{{\cal W}^l_{\bf k} - {\cal W}^l_{{\bf k} - {\bf q}} +
{\cal W}^l_{{\bf k}_1} - {\cal W}^l_{{\bf k}_1 + {\bf q}} \}.
\label{eq:r7}
\end{equation}
Here, we goes over from the function
${\it w}({\bf k},{\bf k}_1; {\bf k}_2,{\bf k}_3)$ (\ref{eq:f6})
to a new function ${\it w}({\bf k},{\bf k}_1;{\bf q})$
\[
{\it w}({\bf k},{\bf k}_1; {\bf k}_2,{\bf k}_3) \vert_{{\bf k}_2 =
{\bf k} - {\bf q}, \, {\bf k}_3 = {\bf k}_1 + {\bf q}}
\equiv {\it w}({\bf k},{\bf k}_1;{\bf q}).
\]
Based on the exact form of the r.h.s. of equation (\ref{eq:r7}),
we define the lifetimes of the plasmon of momentum ${\bf k}$ as follows
\[
\frac{1}{{\tau}_{pl}({\bf k})} =
\int \frac{{\rm d}{\bf k}_1}{(2 \pi)^3}
\int \frac{{\rm d}{\bf q}}{(2 \pi)^3} \,
\frac{N_{eq}^l({\bf k}_1)(N_{eq}^l({\bf k}-{\bf q}) + 1)
(N_{eq}^l({\bf k}_1+{\bf q}) + 1)}
{(N_{eq}^l({\bf k}) + 1)}
\]
\begin{equation}
\times {\it w}({\bf k},{\bf k}_1;{\bf q})
\, (2 \pi) \delta(\omega^l_{\bf k} - \omega^l_{{\bf k}-{\bf q}} +
\omega^l_{{\bf k}_1} - \omega^l_{{\bf k}_1 + {\bf q}}).
\label{eq:t7}
\end{equation}
Here, the integrand has a more involved structure in comparison with a similar
expression for the case of the lifetimes of the hard transverse gluon [34].
The reason of this fact is that the momentum of the soft quasiparticles (plasmons)
becomes of
the same order as momentum transfers and under these conditions one must
take into account the non-trivial character of the frequency dependence $\omega$
of the collective gluon modes on momentum ${\bf k}$
and vertex corrections.

It is not difficult to estimate the order of the expression (\ref{eq:t7}) at
the momentum scale $gT$. Considering the plasmon gas in thermal
equilibrium with hard particles from the heat bath, i.e.
$T^{\ast} \simeq T$, and using the definition of ${\it w}$ function,
we obtain
\begin{equation}
\frac{1}{{\tau}_{pl}} \sim g^3 N_c^2T.
\label{eq:y7}
\end{equation}
However, obtaining a numerical factor of proportionality in (\ref{eq:y7})
is a complicated problem even for limiting case of ${\bf k}=0$-mode.
Here, we restrict our consideration to following general remark,
which somewhat simplifies the matter.

In proving the gauge invariance of the nonlinear Landau damping rate, we have
shown [10] that the $\,^{\ast} \tilde{\Gamma}$ function (\ref{eq:i6}), (\ref{eq:e5})
entering into the definition of the scattering probability (\ref{eq:f6})
can be introduced in its simplest form
\[
\,^{\ast} \tilde{\Gamma}(k,-k_2,k_1,-k_3) =
\]
\[
k^2 k_1^2 k_2^2 k_3^2 \{
\,^{\ast} \Gamma^{0000}(k,-k_2,k_1,-k_3) -
\,^{\ast}{\cal D}_{\rho \alpha}(k - k_3) \,^{\ast} \Gamma^{00 \rho}(k,-k_3,k_1 - k_2)
\,^{\ast} \Gamma^{\alpha 00}(-k_1+k_2,k_1, - k_2)
\]
\[
- \,^{\ast}{\cal D}_{\rho \alpha}(k - k_2) \,^{\ast} \Gamma^{00 \rho}(k,-k_2,k_1 - k_3)
\,^{\ast} \Gamma^{\alpha 00}(-k_1 + k_3,k_1,-k_3) \} \vert_{on-shell}.
\]
However, even if the last expression is accounted for,
after performing the integration over solid angles in $\delta \Gamma^{0000}$
and $\delta \Gamma^{00{\rho}}$ HTL-amplitudes (see, e.g., [28]), we obtain expressions which
are very cumbersome. For this reason passage to the limit $\vert {\bf k} \vert
\rightarrow 0$ is non-trivial.

The estimate (\ref{eq:y7}) shows, that at the momentum scale $gT$ a "collision"
damping rate of the soft longitudinal mode is suppressed by a power of $g$
relative to a value of the nonlinear Landau damping rate ($\sim g^2 T$) [10].
Therefore at the soft momentum scale it can be neglected by the influence
of the plasmon interactions among themselves on the relaxation dynamics of
a plasma excitations.

As was mentioned in Sec. 2, the process of the nonlinear Landau damping
leads to an effective pumping of energy across the spectrum towards small
wavenumbers. By the effect of pumping, all plasmons will tend to
concentrate near small $\vert {\bf k} \vert = \vert {\bf k}_0
\vert \rightarrow 0$. However, phase space, within which the plasmons are occupied,
proportional to $\vert {\bf k}_0 \vert^3$, will also be very small
(a similar state, when a great many plasmons with small wavenumbers
are mainly concentrated in a small
phase volume, in the theory
of strong turbulent electron-ion plasma this is sometimes called {\it a plasmon
condensate} [36]). By virtue of this
fact intensive collisions between plasmons arise and this scattering
process is described by the Boltzmann equation (\ref{eq:n1}).
This leads to a "throw out" of plasmons from region of small
$\vert {\bf k} \vert$ and thus a suppression of the increase of the ${\bf k} = 0$-mode.
In mathematical language this denotes that for definite values of the momentum,
the magnitude of the
off-equilibrium fluctuation $\delta N_{\bf k}^l$ becomes as large as
$N_{eq}^l ({\bf k})$, and therefore a linearization of the Boltzmann  equation
(\ref{eq:n1}) is no longer valid, like the estimation (\ref{eq:y7}), corrected
within the framework of this approximation. In the region $\vert {\bf k} \vert \ll
gT$, where collisions among plasmons start to play a role, it is necessary to solve
the exact nonlinear integro-differential equation, whose r.h.s. is considered as
both the process of the scattering plasmons off thermal particles,
and the process of four-plasmon decays, i.e.
\begin{equation}
\frac{\partial N_{\bf k}^{l}}{\partial t} +
{\bf V}_{\bf k}^{l} \, \frac{\partial N_{\bf k}^{l}}
{\partial {\bf x}} = - 3(\omega_{pl}/g)^2
\int \frac{{\rm d}{\bf k}_{1}}{(2 \pi)^3}
\; ( \omega_{\bf k}^{l} - \omega_{{\bf k}_
{1}}^{l}) \, {\rm Q}({\bf k},{\bf k}_{1})
N_{\bf k}^l N_{{\bf k}_{1}}^{l}
\label{eq:u7}
\end{equation}
\[
+ \int \frac{{\rm d} {\bf k}_1}{(2 \pi)^3} \frac{{\rm d} {\bf k}_2}{(2 \pi)^3}
\frac{{\rm d} {\bf k}_3}{(2 \pi)^3} \,
(2 \pi)^4 \delta (\omega_{\bf k}^l + \omega_{{\bf k}_1}^l - \omega_{{\bf k}_2}^l -
\omega_{{\bf k}_3}^l)
\delta ({\bf k} + {\bf k}_1 - {\bf k}_2 - {\bf k}_3)
\]
\[
\times w({\bf k}, {\bf k}_1; {\bf k}_2, {\bf k}_3)
\, (N_{{\bf k}_1}^l N_{{\bf k}_2}^l N_{{\bf k}_3}^l +
N_{\bf k}^l N_{{\bf k}_2}^l N_{{\bf k}_3}^l -
N_{\bf k}^l N_{{\bf k}_1}^l N_{{\bf k}_2}^l -
N_{\bf k}^l N_{{\bf k}_1}^l N_{{\bf k}_3}^l),
\]
where
\[
{\rm Q}({\bf k}, {\bf k}_{1}) =
(2 \pi)^4 N_{c}
\bigg( \frac{\partial {\rm Re} \, \varepsilon^{l}(k)}{\partial \omega}
\bigg)_{\omega = \omega_{\bf k}^{l}}^{-1}
\bigg( \frac{\partial {\rm Re} \, \varepsilon^{l}(k_{1})}{\partial \omega_{1}}
\bigg)_{\omega_{1} = \omega_{\bf k_1}^{l}}^{-1}
\]
\[
\times \int \, \frac{{\rm d} \Omega}{4 \pi} \delta ( \omega_{\bf k}^{l}
- \omega_{{\bf k}_{1}}^l -
{\bf v}( {\bf k} - {\bf k}_{1}))
\, \vert {\cal M}^{c}({\bf k},{\bf k}_{1})
+ {\cal M}^{\parallel}({\bf k},{\bf k}_{1})
+ {\cal M}^{\perp}({\bf k},{\bf k}_{1}) \vert^{2}.
\]
The amplitudes ${\cal M}^c, \, {\cal M}^{\parallel}$ and ${\cal M}^{\perp}$
have the following forms [10]
\[
{\cal M}^{c}({\bf k},{\bf k}_{1}) =
\frac{g^2}{\vert {\bf k} \vert \vert {\bf k}_{1} \vert} \,
\frac{1}{\omega^l_{\bf k} \omega^l_{{\bf k}_1}}
\, \frac{({\bf k}{\bf v})({{\bf k}_{1}}{\bf v})}
{\omega_{\bf k}^{l} - ({\bf k}{\bf v})},
\]
\[
{\cal M}^{\parallel}({\bf k},{\bf k}_{1})
= \frac{g^2}{\vert {\bf k} \vert \, \vert {\bf k}_{1} \vert} \,
\frac{(({\bf k}-{\bf k}_1) {\bf v})}{({\bf k} - {\bf k}_1)^2}
\]
\[
\times \left( \frac{(k - k_1)^2 \,^{\ast}\Delta^l(k - k_1)}{\omega \omega_1 (\omega - \omega_1)^2}
\, \delta \Gamma^{ijl}(k,-k_1,-k + k_1)k^ik^j_1(k - k_1)^l \right)_{\omega
= \omega_{\bf k}^{l}, \; \omega_1 = \omega_{{\bf k}_{1}}^{l}},
\]
\[
{\cal M}^{\perp}({\bf k},{\bf k}_1) =
\frac{g^2}{ \vert {\bf k} \vert \vert {\bf k}_1 \vert}
\frac{([{\bf n}({\bf k} - {\bf k}_1)]{\bf v})}{{\bf n}^2({\bf k} - {\bf k}_1)^2}
\]
\[
\times \left( \frac{\,^{\ast}\Delta^t(k - k_1)}{\omega \omega_1}
\,^{\ast} \Gamma^{ijl}(k,-k_1,-k + k_1)
k^ik^j_1 [{\bf n}({\bf k} - {\bf k}_1)]^l \right)_{\omega=
\omega^l_{\bf k}, \; \omega_1= \omega^l_{{\bf k}_1}}.
\]
Here, ${\bf n} \equiv [{\bf k}{\bf k}_1]$. We present the amplitudes
${\cal M}^c, \,{\cal M}^{\parallel}$ and ${\cal M}^{\perp}$ in their simplest
form, defined by temporal gauge.

A qualitative analysis of the solution behavior of a similar equation in the case of
electron-ion
plasma with a rather crude approximation of the kernel ${\rm Q}({\bf k},{\bf k}_1)$
and the probability of plasmon-plasmon scattering
${\it w}({\bf k},{\bf k}_1;{\bf k}_2, {\bf k}_3)$, was performed
by Kovrizhnykh [30]. He has shown, in particular, that really in the
kinematic regime of small wavenumbers $\vert {\bf k} \vert$, the process of
plasmon-plasmon scattering starts to play a dominant role, that prevents
an increase of the ${\bf k} = 0$-mode. In the case of the the non-Abelian plasma, an analogous
qualitative investigation of the solution of equation (\ref{eq:u7}) presents difficulties
because of the above-mentioned complexity of the integrands on the r.h.s. of
equation (\ref{eq:u7}), and therefore it requires some additional approximations,
allowing these functions to be made more visible and suitable for numerical
calculations.

\section{\bf Conclusion}
\setcounter{equation}{0}

Within the framework of the semiclassical kinetic theory of a hot gluon plasma
we have obtained the transport equation for the plasmons, taking into account
four-plasmon decay. The probability of plasmon-plasmon scattering at the leading order in the
coupling constant is derived. This is defined with the help of three-gluon and
four-gluon effective vertices, and the effective propagator only, as the probability
of the process of scattering of the plasmon by hard QGP particles [10].
It is proved that the scattering probability is gauge-independent at least within a class of
covariant and temporal gauges.

In this paper we have restricted our consideration to a derivation of the Boltzmann
equation for the simplest collective Bose-modes of hot gluon plasma, colorless
plasmons. This fact was reflected in deciding on
a trivial color structure of the plasmon occupation number:
$N_{\bf k}^{l \, ab} \sim \delta^{ab}$. It is clear that such a choice
of trivial color structure is completely neglected by the fundamental property
of plasmons in the non-Abelian plasma, such as the availability of color charge.
This is a very important difference from the Abelian case, where the plasmons
do not carry electric charge. While the initial state of the problem can have a
trivial color structure, the dynamics of these collective modes can change
the color of the plasmons. In light of these remarks, here
we are dealing with an Abelianized version of more complicated colored plasmon
dynamics. It is a first step towards a full description of plasmon transport
properties to derive the Boltzmann equation
which takes into account the precession of the color charge of the plasmon.
It is probably only in the construction of such a complete equation that we
gain a reasonable explanation of the structure of the scattering probability (\ref{eq:f6}).

The scheme of the derivation of the Boltzmann equation for colorless plasmons, proposed
in this paper, admits the straightforward generalization to the case, when the
number density of the plasmon
$N_{\bf k}^{l \, ab}$ has a non-trivial color structure, e.g. such as
$N_{\bf k}^{l \, ab} = (T^c)^{ab} N_{\bf k}^{l \, c}$.
In this case,
as was discussed in Sec. 4, it is necessary to assume that the regular (background) part of the
field $A_{\mu}^R(X)$ is different from zero. The density matrix, effective
propagator, three-gluon, four-gluon vertices become the functionals of the
background field $A_{\mu}^R (X)$ [9, 18], which assumed vanishing at
$X_0 \rightarrow - \infty$. This significantly complicates the problem of
construction of the required kinetic equation.
It can somewhat  simplify the problem if it is considered
that the fluctuation $\delta N_{\bf k}^{l \, ab}$ is a small perturbation in relation
to the equilibrium value $\delta^{ab} N_{eq}^l ({\bf k})$, where
$N_{eq}^l ({\bf k})$ is the Planck distribution, and restrict the consideration
to the linearized Boltzmann equation for colored plasmons. The exact consideration
of a given question will be subject of separate research.

There is an independent test of the validity of the derived Boltzmann equation
(\ref{eq:n1}) (exactly, its linearized version (\ref{eq:r7})).
As was mentioned in introduction in [18] the fundamental derivation of the Boltzmann equation for a high temperature
Yang-Mills plasma was proposed by Blaizot and Iancu within the
CTP formalism framework. Their derivation relies on a gauge-covariant
gradient expansion of the Kadanoff-Baym equations for the gluon two-point
function. The Boltzmann equation has emerged as the quantum transport equation
at leading order in $g$ for the gauge-covariant fluctuation
$\delta \acute{G}$
of a hard gluon propagator.

Besides the above-mentioned paper, the Kadanoff-Baym equations for the
off-equilibrium propagator of the soft gluon ${\cal D}_{\mu \nu} (X, Y)$,
which are formally identical to those for a hard gluon propagator
$G_{\mu \nu} (X, Y)$, are written out. These equations are used in [18]
only to deduce the relation between the off-equilibrium gauge-covariant
fluctuation
$\acute{{\cal D}^{<}} (k, x)$ and the gauge-covariant
fluctuation of the leading-order soft gluon polarization
tensor $\acute{\delta \Pi^{<}} (k, x)$, and the problem of self-interactions of
the soft fields is not considered here. However, in principal there is nothing
to forbid the use of these equations for research in the dynamics of the soft
fields and construction of the relevant transport equation within the framework of the scheme,
suggested by Blaisot and Iancu [18].
Here, by $\acute{\delta \Pi^{<}} (k,x)$ we mean the fluctuation of the
next-to-leading order of the soft gluon self-energy, involving three- and
four-gluon off-equilibrium vertices with soft external lines. The relevant
effects of self-interaction of the soft fields are encoded in these functions.
Here, we note that in the real-time formalism, technical complications
resulting from the doubling of degrees of freedom are arisen. However in recent
years a suitable technique allowing one effectively to work not only with the single-particle
propagator, but also with three- and four-point functions, has been developed
[8]. This greatly simplifies calculations in real time and enables us to derive
the transport equation for the soft gluons, in particular, the plasmons, directly
from the underlying quantum field theory and compare it with the equation, obtained
in this paper in the context of the semiclassical approximation.
This formal scheme is also needed to specify
the limits of validity of the semiclassical kinetic approach
to the research of the processes of nonlinear interaction of the soft fields in a hot QCD plasma.

\section*{\bf Acknowledgement}

We are very grateful to Stanislaw Mr\'owczy\'nski for useful correspondence.
This work was supported in part by Grant INTAS (No.\,2000-15) and
Grant for Young Scientist of Russian Academy of Sciences (No.\,1999-80).


\begin{thebibliography}{}
\bibitem{in}
In our paper (Markov and Markova in Ref. [13]) it was shown that existence
of the spin degree of freedom for gluons in the semiclassical approximation
results in two covariant coupled dynamical equations for the distribution
function $f$ and the function related to circular polarization.
Therefore if required, more subtle effects associated with gluons polarization
can be without difficulty taken into account in the derivation scheme of kinetic
equation that outlined at present work.
\end{thebibliography}
\end{document}